\documentclass[a4paper,12pt]{article}
\usepackage{epsfig}
\usepackage{amsmath}
\usepackage{amssymb}
\usepackage{amsfonts}
\usepackage{bbm}
\usepackage{fleqn}
\usepackage[footnotesize]{caption}
\usepackage{graphicx}
\usepackage{mathrsfs}
\usepackage[center,footnotesize,hang]{subfigure}

\def\simlt{\stackrel{<}{{}_\sim}}

\def\be{\begin{equation}}
\def\ee{\end{equation}}
\def\beq{\begin{equation}}
\def\eeq{\end{equation}}
\def\bea{\begin{eqnarray}}
\def\eea{\end{eqnarray}}

\def\<{\left\langle}
\def\>{\right\rangle}

\def\simlt{\stackrel{<}{{}_\sim}}

\def\be{\begin{equation}}
\def\ee{\end{equation}}
\def\beq{\begin{equation}}
\def\eeq{\end{equation}}
\def\bea{\begin{eqnarray}}
\def\eea{\end{eqnarray}}

\newcommand{\newc}{\newcommand}
\newc{\ol}{\overline}
\newc{\wt}{\widetilde}
\newc{\bs}{\boldsymbol}
\newc{\ma}{\mathcal}
\newc{\vl}{\langle}
\newc{\vr}{\rangle}
\def\vev#1{\langle #1 \rangle}
\newc{\sg}{S}
\newc{\ug}{U}
\newc{\tg}{T}

\allowdisplaybreaks[2]
\begin{document}
\bibliographystyle{OurBibTeX}

\title{\hfill ~\\[-30mm]
                  \textbf{Minimal predictive see-saw model with normal neutrino mass hierarchy}        }
\date{}
\author{\\[-5mm]
Stephen F. King\footnote{E-mail: {\tt king@soton.ac.uk}}\\ \\
  \emph{\small School of Physics and Astronomy, University of Southampton,}\\
  \emph{\small Southampton, SO17 1BJ, United Kingdom}\\[4mm]}

\maketitle

\begin{abstract}
\noindent
{We consider the type I see-saw model with two right-handed neutrinos and a normal neutrino mass hierarchy 
and impose a zero coupling between 
the right-handed neutrino mainly responsible for the atmospheric neutrino mass and the electron neutrino. We derive a master formula which relates see-saw input parameters 
in a one to one correspondence
with physical neutrino observables. Using the master formula we search for 
simple ratios of couplings consistent with current data on neutrino mass and lepton mixing.
We discover a minimal predictive example in which the right-handed neutrino mainly responsible for the atmospheric neutrino mass has couplings to $(\nu_e, \nu_{\mu}, \nu_{\tau})$ proportional to $(0,1,1)$ and the 
right-handed neutrino mainly responsible for the solar neutrino mass has couplings to $(\nu_e, \nu_{\mu}, \nu_{\tau})$ proportional to $(1,1,3)$ or $(1,3,1)$, with a
relative phase $\eta = \mp \pi/3$, providing the link between leptogenesis and CP violation in neutrino oscillation experiments.
We show how these patterns of couplings could arise from an $A_4$ family symmetry
model of leptons which predicts all the PMNS parameters in terms of the neutrino mass ratio
$m_2/m_3$, 
corresponding to Tri-bimaximal-Cabibbo mixing, 
accurate to one degree, with the prediction $\delta \approx \pm \pi/2$. } 
 \end{abstract}
\thispagestyle{empty}
\vfill
\newpage
\setcounter{page}{1}

\section{Introduction}

The type I see-saw mechanism \cite{Minkowski:1977sc}
provides a beautiful understanding of the smallness of neutrino masses
as being due to the heavy right-handed Majorana neutrino masses.
Furthermore it strongly suggests that light neutrinos are also Majorana.
However, despite its attractive features, the see-saw mechanism provides 
no understanding of the observed bi-large lepton mixing.
It also provides no insight into 
either the ordering (i.e. normal or inverted) or the mass scale of the neutrinos (i.e. the mass of the lightest neutrino).
Moreover the see-saw mechanism is difficult to test experimentally, unless the right-handed neutrino masses are at the TeV scale \cite{King:2004cx}, and
typically contains more parameters than physical observables \cite{Davidson:2004wi}. 
The larger number of see-saw parameters
means that the neutrino Yukawa couplings cannot ever be determined from physical neutrino observables. 
Finally, apparently ``unnatural'' cancellations can take place when constructing the effective neutrino mass matrix from the see-saw parameters.

One attractive idea which avoids the last problem of ``unnatural'' cancellations in the see-saw mechanism
is that the right-handed neutrinos contribute sequentially to the light effective neutrino mass matrix
with hierarchical strength, leading to a normal mass hierarchy of physical neutrino masses $m_3\gg m_2\gg m_1$.
This is consistent with recent {\it Planck} results which provide no evidence of quasi-degeneracy and lead to a bound 
\cite{Ade:2013lta},
\be
\sum_i m_i < 0.23 {\rm eV} \  (95\%; Planck+WP+highL+BAO).
\ee
The idea of such a sequential dominance (SD) \cite{King:1998jw} of right-handed neutrinos is that one right-handed neutrino of mass $M_{\rm atm}$ dominantly contributes to the see-saw mechanism
and is mainly responsible for the atmospheric neutrino mass $m_3$, while a second sub-dominant right-handed neutrino of mass $M_{\rm sol}$ is mainly responsible for the solar neutrino mass $m_2$, with a third almost decoupled right-handed neutrino of mass $M_{\rm dec}$ 
being mainly responsible for the lightest neutrino mass $m_1$.
Moreover, in the limit that the lightest neutrino mass $m_1\rightarrow 0$,
the almost decoupled right-handed neutrino becomes irrelevant and the see-saw mechanism reduces to the two right-handed neutrino case with a normal neutrino mass hierarchy, which reduces the number of see-saw parameters
as we now discuss.

The type I see-saw mechanism with effectively two right-handed neutrinos \cite{King:1999mb}
has fewer parameters and predicts one massless neutrino, leading to simplified formulas for neutrinoless double beta decay and the sum of neutrino masses relevant for cosmology. 
Texture zeros can further reduce the number of see-saw parameters
\cite{King:2002nf,Frampton:2002qc,Raidal:2002xf,King:2002qh,Ibarra:2003up}.
With one texture zero \cite{King:2002nf,King:2002qh}, the $3\times 2$ neutrino Yukawa matrix 
contains 7 real parameters, after charged lepton phase rotations, which is exactly the same number of parameters as non-zero physical neutrino observables, 
comprising 2 real positive masses $m_2,m_3$, the 3 mixing angles $\theta_{23}, \theta_{13}, \theta_{12}$, the oscillation phase $\delta$ and the single Majorana phase $\beta$.

In the case of the two right-handed neutrino model with a normal hierarchy,
if the ``dominant'' right-handed 
neutrino has zero coupling to the lepton doublet containing the electron, 
this was shown over a decade ago to bound the reactor angle $\theta_{13} \simlt m_2/m_3$ 
\cite{King:2002nf,King:2002qh}. We shall refer to this zero coupling as a ``dominant texture zero''.
For example, such a ``dominant texture zero'' could appear in the (1,1) entry of the neutrino Yukawa matrix,
similar to the quark and charged lepton Yukawa matrices, where a (1,1) texture zero
it is a key ingredient of the Gatto-Sartori-Tonin (GST)
\cite{Gatto:1968ss} and Georgi-Jarlskog (GJ) relations \cite{Georgi:1979df}.
Recently, Daya Bay  \cite{An:2012eh} and RENO \cite{Ahn:2012nd} have measured $\theta_{13}\approx 0.15$ 
near the upper limit of the bound $\theta_{13} \simlt m_2/m_3$ \cite{King:2002nf,King:2002qh}.
The fact that this bound is saturated could be indicative of a particular underlying pattern of Yukawa couplings, and this provides
part of the motivation for revisiting the minimal see-saw model. In particular we shall derive a new ``master formula''
which relates physical observables to Yukawa couplings, providing the necessary tools for such questions
to be addressed. 

The measurement of a non-zero reactor angle excludes tri-bimaximal (TB) mixing \cite{Harrison:2002er}.
It is nevertheless convenient to express the $s$olar,
$a$tmospheric and $r$eactor angles in terms of (small) deviation parameters ($s$, $a$ and
$r$) from TB mixing~\cite{King:2007pr} (for a related parametrisation see \cite{Pakvasa:2007zj}): 
\be
\label{rsadef}
\sin \theta_{12}=\frac{1}{\sqrt{3}}(1+s),\ \ \ \ 
\sin\theta_{23}=\frac{1}{\sqrt{2}}(1+a), \ \ \ \ 
\sin \theta_{13}=\frac{r}{\sqrt{2}}.
\ee
For example, following the Daya Bay and RENO results, it was pointed out that 
the lepton mixing angles are consistent with the so called Tri-bimaximal-Cabibbo 
(TBC) ansatz~\cite{King:2012vj} where 
\beq
s=0, \ \ \ \ a=0, \ \ \ \ r=\lambda ,
\label{ansatzTBC}
\eeq
corresponding to 
\beq
\sin \theta_{12}=\frac{1}{\sqrt{3}}, \ \ \ \ \sin \theta_{23}=\frac{1}{\sqrt{2}}, \ \ \ \ \sin \theta_{13}=\frac{\lambda}{\sqrt{2}},
\label{ansatz}
\eeq
with $\lambda = 0.225$ being the Wolfenstein parameter, yielding the TBC angles,
\beq
\theta_{12}=35.26^{\circ}, \ \ \ \ \theta_{23}=45^{\circ},\ \ \ \   \theta_{13}=9.15^{\circ}. 
\label{TBCangles}
\eeq
The TBC lepton
mixing matrix is given to second order by~\cite{King:2012vj}, 
\beq
V_{\mathrm{TBC}} \approx
\left( \begin{array}{ccc}
\sqrt{\frac{2}{3}}(1-\frac{1}{4}\lambda^2)  & \frac{1}{\sqrt{3}}(1-\frac{1}{4}\lambda^2) 
& \frac{1}{\sqrt{2}}\lambda e^{-i\delta } \\
-\frac{1}{\sqrt{6}}(1+\lambda e^{i\delta })  & \frac{1}{\sqrt{3}}(1- \frac{1}{2}\lambda e^{i\delta })
& \frac{1}{\sqrt{2}}(1-\frac{1}{4}\lambda^2) \\
\frac{1}{\sqrt{6}}(1- \lambda e^{i\delta })  & -\frac{1}{\sqrt{3}}(1+ \frac{1}{2}\lambda e^{i\delta })
 & \frac{1}{\sqrt{2}}(1-\frac{1}{4}\lambda^2)
\end{array}
\right)+ \mathcal{O}(\lambda^3).
\label{TBC}
\eeq
TBC mixing thus assumes maximal atmospheric mixing and trimaximal solar mixing.
For the normal neutrino mass hierarchy, some of the global fits, reviewed in \cite{King:2013eh}, prefer an
alternative ansatz which we refer to as TBC2, corresponding to,
\beq
s=-\lambda^2, \ \ \ \ a=-\lambda /2, \ \ \ \ r=\lambda .
\label{ansatzTBC2}
\eeq
This TBC2 ansatz has the feature that the atmospheric mixing angle is in the first octant and the solar mixing angle
is somewhat less than its tri-maximal value as preferred by some of 
the latest global fits at the one sigma level, while TBC mixing generally 
remains valid at the three sigma level or better for all the global fits \cite{King:2013eh}.
We emphasise that the issue of whether the atmospheric angle is maximal or not remains
unresolved, with SuperKamiokande having a mild preference 
for a non-maximal atmospheric angle \cite{SK} while T2K prefers a maximal atmospheric angle,
although both preferences are less than one sigma and hence not statistically significant.
This situation is expected to be clarified in the near future.
Returning to our analysis, the ansatze we have used will serve as an approximate guide
in searching for particular patterns of Yukawa couplings. Once a particular pattern is identified,
it then takes a life of its own, independently of the TBC or TBC2 ansatz, and leads to predictions for lepton
mixing independently of either of these two ansatze which were used to discover the pattern in the first place.

The present paper divides into two parts.
In the first part we shall perform a model independent analysis
of the two right-handed neutrino model with a normal neutrino mass
hierarchy and a single ``dominant texture zero'' in the electron component of the 
Yukawa matrix \cite{King:2002nf,King:2002qh}.
This model provides an attractive and rather 
minimal extension of the Standard Model consistent with all current neutrino data,
with a one to one correspondence between the Yukawa couplings and the physical observables.
We shall derive a master formula comprising
exact analytic results which express the Yukawa couplings in terms of general values of 
the physical masses and mixings. These results in principle do not rely on the SD assumption
and are valid even for large values of reactor angle.
However we show that, in practice, the observed
reactor angle $\theta_{13} \simlt m_2/m_3$ implies and is implied by SD, where 
the ``texture zero'' in the electron component becomes the ``dominant texture zero''.
We also confirm that a second texture zero is not possible, at least
without invoking charged lepton corrections which we do not consider in this paper.
In fact the two right-handed neutrino model with two texture zeros
\cite{Frampton:2002qc,Raidal:2002xf,Ibarra:2003up} was recently shown to be not viable 
for the normal hierarchy case \cite{Harigaya:2012bw,Shimizu:2012ry}, which provides a further reason why 
we mainly focus on the one texture zero case \cite{King:2002nf,King:2002qh,Ibarra:2003up} here. 

The second part of the paper is concerned with applying the results obtained in the first part to model
building. An important step in this direction is to use the master formula to derive
exact analytic results for the ratios of Yukawa couplings in terms of physical neutrino observables, 
and to use these results, together with the current best fit values of neutrino masses and mixing angles, to 
search for simple patterns of Yukawa couplings. We find that 
the atmospheric neutrino mass can have 
couplings to $(\nu_e, \nu_{\mu}, \nu_{\tau})$ proportional to, for example, $(0,1,1)$ or $(0,1,2)$,
which are strongly dependent on the value of the CP violating oscillation phase $\delta$.
The right-handed neutrino mainly responsible for the solar neutrino mass can have 
couplings to $(\nu_e, \nu_{\mu}, \nu_{\tau})$ proportional to $(1,4,2)$ or $(1,1,3)$ or $(1,3,1)$,
which are strongly dependent on the value of both the oscillation phase and the Majorana phase
and differ from the traditional constrained sequential dominance (CSD) values 
\cite{King:2005bj}. In particular we focus on CSD3 defined to be based on the atmospheric
couplings $(0,1,1)$ and the solar couplings being either $(1,1,3)$ or $(1,3,1)$, 
with a relative phase difference $\eta = \mp \pi/3$.
We show how these patterns of couplings could arise from an $A_4$ family symmetry
model of leptons which predicts all the PMNS parameters in terms of the neutrino mass ratio
$m_2/m_3$, 
corresponding to approximate Tri-bimaximal-Cabibbo mixing 
to an accuracy of one degree with $\delta \approx \pm \pi/2$.
The model provides a link between leptogenesis and CP violation in neutrino oscillation experiments.

We remark that, although the two right-handed neutrino model with one texture zero case 
has been discussed before \cite{King:2002nf,King:2002qh,Ibarra:2003up}, such studies
were performed about a decade ago when the mixing angles were not known so precisely,
so it is timely to revisit them,
in the light of the first measurement of the reactor angle by Daya Bay and RENO which approximately 
saturates the ``dominant texture zero'' bound $\theta_{13} \simlt m_2/m_3$ \cite{King:2002nf,King:2002qh}.
Moreover, as indicated above, we present new and 
exact analytic results which directly relate Yukawa couplings to physical neutrino observables.
Previously in \cite{King:2002nf,King:2002qh} only approximate relations were given,
while in \cite{Ibarra:2003up} exact relations between Yukawa couplings and physical neutrino observables
were only given parametrically via the complex angle $z$ arising from an orthogonal parametrisation.
Here we give a simple derivation of exact analytic relations between Yukawa couplings and physical neutrino observables via a master formula
which does not rely on any parameterisation.
We also emphasise a further important difference between the present paper and the earlier ones
is that we know now that the bound $\theta_{13} \simlt m_2/m_3$ is approximately saturated, and in this
paper we seek a simple answer to this question in terms of ratios of Yukawa couplings with simple
phase relations. The fact that we find a simple answer, namely CSD3,
is very gratifying, since it opens up new possibilities for model building as well as linking
leptogenesis to the PMNS phases. We go on to construct an explicit $A_4$ model of leptons 
for one of the examples that completely determines the PMNS mixing matrix including all the phases.

The layout of the remainder of this paper is as follows.
In the first part of the paper we perform a model independent analysis.
In section~\ref{SD} we briefly show how the two right-handed neutrino model emerges as the limiting case of 
a three right-handed neutrino model with sequential dominance. 
In section~\ref{2RH} we discuss the two right-handed neutrino with a normal neutrino mass hierarchy and a texture zero in the electron component of either one of the two right-handed neutrinos,
and derive a master formula relating see-saw parameters to physical parameters for this case.
Using the master formula we show that SD emerges
as a consequence of bi-large mixing and the observed reactor angle,
with the texture zero being associated with the dominant right-handed neutrino.
In section~\ref{search}, using the master formula,
we calculate ratios of Yukawa couplings and see-saw phases, as functions of the PMNS phases,
for particular mixing patterns such as TBC and TBC2 which are consistent with present data.

In the second part of the paper we discuss the implications of the results for model building.
After a brief review in section~\ref{models} of the status of indirect models, in section~\ref{models2}
we scrutinise the results obtained in the previous
section, looking for simple possibilities which could form the basis of new indirect models.
We discuss three such examples in some detail, checking
their viability using exact numerical results for PMNS parameters and
leading order approximate analytic results in section~\ref{LO}.
We also discuss the link between leptogenesis and PMNS phases in section~\ref{LG}.
In section~\ref{CSD3} we then focus on CSD3 based on the alignment  $(0,1,1)$
for the atmospheric
column and $(1,1,3)$ or $(1,3,1)$ for the solar column with a relative phase difference $\eta = \mp \pi/3$ and 
show how these follow from $A_4$ family symmetry model of leptons 
based on F-term vacuum alignment and spontaneous CP violation.
The model predicts all the PMNS mixing parameters in terms of the neutrino mass ratio,
corresponding to approximate Tri-bimaximal-Cabibbo mixing 
to an accuracy of one degree with $\delta \approx \pm \pi/2$.
Section~\ref{conclusion} concludes the paper.


\section{Two Right-Handed Neutrino Model as a Limiting Case of Sequential Dominance}
\label{SD}

The starting point for our analysis is the see-saw mechanism in the flavour basis
where the charged lepton mass matrix $M_E$ is diagonal with real positive eigenvalues
$m_e, m_{\mu}, m_{\tau}$ and the three right-handed neutrino
Majorana mass matrix $M_R$ is also diagonal, with real positive eigenvalues,
$M_{\rm atm}, M_{\rm sol}, M_{\rm dec}$,
\begin{equation}
M_{E}=
\left( \begin{array}{ccc}
m_{ e} & 0 & 0    \\
0 & m_{ \mu} & 0 \\
0 & 0 & m_{ \tau}
\end{array}
\right), \
M_{R}=
\left( \begin{array}{ccc}
M_{\rm atm} & 0 & 0    \\
0 & M_{\rm sol} & 0 \\
0 & 0 & [M_{\rm dec}]
\end{array}
\right).
\label{seq1}
\end{equation}
We write the neutrino Dirac mass matrix as
\begin{equation}
m^D=
\left( \begin{array}{ccc}
m^D_{e, \rm atm} & m^D_{e, \rm sol} & [m^D_{e, \rm dec}]   \\
m^D_{\mu, \rm atm} & m^D_{\mu, \rm sol} & [m^D_{\mu, \rm dec}] \\
m^D_{\tau, \rm atm} & m^D_{\tau, \rm sol} & [m^D_{\tau, \rm dec}]
\end{array}
\right)
\equiv \left( \begin{array}{ccc}
m^D_{\rm atm} & m^D_{\rm sol} & [m^D_{\rm dec}]  
\end{array}
\right),
\label{dirac}
\end{equation}
in the convention where the effective Lagrangian after electroweak symmetry breaking, 
with the Higgs vacuum expectation value (vev) inserted, 
is given by
\beq
{\cal L}=-  \overline{E_L} M_{E} E_R  -  \overline{\nu_L} m^{D} N_R 
- \frac{1}{2}\overline{N_R^c} M_{R} N_{R} 
+ {H.c.}\; ,
\eeq
where $\nu_L =(\nu_{e},\nu_{\mu }, \nu_{\tau })$ are the three left-handed neutrino fields
which appear together with $E_L=(e_L,\mu_L,\tau_L)$ in the lepton doublets $L=(L_{e},L_{\mu }, L_{\tau })$
and $N_R=(N_{\rm atm},N_{\rm sol}, N_{\rm dec})$ 
are the three right-handed neutrinos and we have defined the three Dirac column vectors as
$m^D_{\rm atm}$, $m^D_{\rm sol}$, $m^D_{\rm dec}$.

The term for the light neutrino masses in the effective Lagrangian (after electroweak symmetry breaking), resulting from integrating out the massive right
handed neutrinos (i.e. the see-saw mechanism
with the light effective neutrino Majorana mass matrix $m^{\nu}=m^DM_R^{-1}{m^D}^T$) is
\begin{equation}
\mathcal{L}^\nu_{eff} =
 \frac{(\overline{\nu_L} m^D_{\rm atm})({m^D_{\rm atm}}^{T} \nu_L^c)}{M_{\rm atm}}
 +\frac{(\overline{\nu_L} m^D_{\rm sol})({m^D_{\rm sol}}^{T} \nu_L^c)}{M_{\rm sol}}
\ \left[ \ +\frac{(\overline{\nu_L} m^D_{\rm dec})({m^D_{\rm dec}}^{T} \nu_L^c)}{M_{\rm dec}} \ \right] \label{leff}.
\end{equation}

Sequential dominance (SD) then corresponds to the third
term being negligible, the second term subdominant and the first term
dominant:
\beq\label{SDcond}
\frac{m^D_{\rm atm}{m^D_{\rm atm}}^T}{M_{\rm atm}} \gg
\frac{m^D_{\rm sol}{m^D_{\rm sol}}^T}{M_{\rm sol}} \ \left[ \ \gg
\frac{m^D_{\rm dec}{m^D_{\rm dec}}^T}{M_{\rm dec}}\ \right] \, ,
\label{SD1}
\eeq
which immediately predicts a normal neutrino mass hierarchy,
\beq
\label{normal2}
m_3 \gg  m_2 \ \left[\  \gg m_1  \ \right],
\eeq
which is the main prediction of SD.

We have labelled the dominant
right-handed neutrino and Yukawa couplings mainly responsible for the atmospheric neutrino mass $m_3$ as ``${\rm atm}$'', the subdominant
ones mainly responsible for the solar neutrino mass $m_2$ as ``${\rm sol}$'', and the almost decoupled (sub-sub-dominant) ones
mainly responsible for $m_1$ as ``${\rm dec}$''. 
Note that the mass ordering of right-handed neutrinos is 
not yet specified. We shall 
order the right-handed neutrino masses as $M_1<M_2<M_3$,
and subsequently identify $M_{\rm atm},M_{\rm sol},M_{\rm dec}$ with $M_1,M_2,M_3$
in all possible ways. 

It is clear that in the limit that $m_1\rightarrow 0$ then 
the sub-sub-dominant right-handed neutrino and its associated couplings labelled by ``${\rm dec}$''
decouple completely and the above model reduces to a two right-handed neutrino model.
In that limit we simply drop the third terms [in square brackets] 
in Eqs.\ref{seq1}-\ref{normal2} in anticipation of this.

\section{The Two Right-Handed Neutrino Model with Normal Hierarchy and Dominant Texture Zero}
\label{2RH}

\subsection{Derivation of the Master Formula}
Without assuming SD, we write the see-saw matrices in a simple notation as,
\begin{equation}
m^D=
\left( \begin{array}{cc}
0 & a   \\
e & b \\
f & c 
\end{array}
\right), \ \ \ \
M_{R}=
\left( \begin{array}{cc}
Y & 0     \\
0 & X 
\end{array}
\right).
\label{2rh1}
\end{equation}
where we have written 
the complex Dirac masses as $a,b,c,d,e,f$ with $d=0$
and the real positive right-handed neutrino masses as $Y,X$.
We are in a basis where the charged lepton mass matrix $M_E$ is diagonal with real positive eigenvalues ordered as 
$m_e, m_{\mu}, m_{\tau}$.
Later on we shall show that the observed
reactor angle $\theta_{13} \simlt m_2/m_3$ implies and is implied by SD, where 
the ``texture zero'' in the electron component becomes the ``dominant texture zero'',
i.e. we will later identify $d=m^D_{e, \rm atm}=0$ 
and $Y=M_{\rm atm}$
in the notation of the previous section. 
However, to begin with, we do not presume this identification. We also allow the real positive right-handed neutrino masses $Y$ and $X$ to have either mass ordering, i.e. namely 
$Y=M_1$ or $Y=M_2$ where $M_1<M_2$. Hence the texture zero in Eq.\ref{2rh1} covers both cases with a texture zero in the electron component,
\begin{equation}
m^D=
\left( \begin{array}{cc}
0 & \times   \\
\times  & \times  \\
\times  & \times  
\end{array}
\right), \ \ \ \
\left( \begin{array}{cc}
\times  & 0   \\
\times  & \times  \\
\times  & \times  
\end{array}
\right).
\label{2rh2}
\end{equation}

We now derive a master formula in which the see-saw parameters in Eq.\ref{2rh1} may be related to the low energy physical observables such as neutrino masses, mixing angles and physical phases.
Integrating out the right-handed neutrinos the see-saw formula gives the approximate light effective Majorana neutrino mass matrix,
\beq
m^{\nu}=-m^DM_{R}^{-1}{m^D}^T\ .
\label{seesaw}
\eeq
in the convention where the effective Lagrangian is given by
\footnote{Note that this convention for the 
light effective Majorana neutrino mass matrix $m^{\nu}$
differs by an overall complex conjugation compared to some other 
conventions in the literature.}
\begin{eqnarray}
{\cal L}=-  \overline{E_L} M_{E} E_R 
- \frac{1}{2}\overline{\nu_L} m^{\nu} \nu_{L}^c 
+ {H.c.}\; .
\label{Leff}
\end{eqnarray}
Performing the transformation from the flavour basis to the real positive mass basis by,
 \begin{eqnarray}\label{eq:DiagMe}
V_{E_L} \, M_{E} \,
V^\dagger_{E_R} =
\mbox{diag}(m_e,m_\mu,m_\tau)
 , \quad~
V_{\nu_L} \,m^{\nu}\,V^T_{\nu_L} =
\mbox{diag}(m_1,m_2,m_3),
\label{mLLnu}
\end{eqnarray}
the PMNS matrix is given by
\begin{eqnarray}\label{Eq:PMNS_Definition}
U
= V_{E_{L}} V^\dagger_{\nu_{L}} \,.
\end{eqnarray}
Since, as before, we are in the basis where the charged lepton mass matrix
$M_E$ is already diagonal, then in general $V_{E_{L}}$ can only be a diagonal matrix,
\beq
V_{E_{L}}=P_{E}=\left( \begin{array}{ccc}
e^{i\phi_{e}} & 0  & 0   \\
0 & e^{i\phi_{\mu}}  & 0 \\
0 & 0 & e^{i\phi_{\tau}}
\end{array}
\right),
\label{PE}
\eeq
consisting of arbitrary phases, where
an identical phase rotation on the right-handed charged leptons 
$V_{E_{R}}=P_{E}$ leaves the diagonal charged lepton masses in $M_E$ unchanged.
In practice the phases in $P_{E}$ are chosen to absorb three phases 
from the unitary matrix 
$V^\dagger_{\nu_{L}}$ and to put $U$ in a standard convention,
\beq
U =V P
\label{DM}
\eeq
where, analogous to the CKM matrix,
\bea
 \label{eq:matrix}
V = 
\left(\begin{array}{ccc}
    c_{12} c_{13}
    & s_{12} c_{13}
    & s_{13} e^{-i\delta}
    \\
    - s_{12} c_{23} - c_{12} s_{13} s_{23} e^{i\delta}
    & \hphantom{+} c_{12} c_{23} - s_{12} s_{13} s_{23}
    e^{i\delta}
    & c_{13} s_{23} \hspace*{5.5mm}
    \\
    \hphantom{+} s_{12} s_{23} - c_{12} s_{13} c_{23} e^{i\delta}
    & - c_{12} s_{23} - s_{12} s_{13} c_{23} e^{i\delta}
    & c_{13} c_{23} \hspace*{5.5mm}
    \end{array}\right),
\eea
and the Majorana phase matrix factor is, 
\beq
P  =
\label{Maj}
\left(\begin{array}{ccc}
e^{i\frac{\beta_1}{2}} & 0 & 0 \\
0 & e^{i\frac{\beta_2}{2}} & 0\\
0 & 0 & 1 \\
\end{array}\right).
\label{MNS}
\eeq
From Eqs.\ref{mLLnu},\ref{Eq:PMNS_Definition},\ref{PE}, we find,
\beq
U^{\dagger}P_E m^{\nu} P_E U^{*}=\mbox{diag}(m_1,m_2,m_3).
\eeq
Then using Eq.\ref{DM} we find
\beq
\label{Maj2}
V^{  \dagger}P_E m^{\nu} P_E V^{ *} =
P \mbox{diag}(m_1,m_2,m_3) P =\mbox{diag}(0,e^{i\beta}m_2,m_3),
\eeq
where we have dropped $m_1=0$ (since we are considering the two right-handed neutrino model with a normal
mass hierarchy) and defined $\beta\equiv \beta_2$.
Using the see-saw formula, Eq.\ref{seesaw}, this may be written,
\beq
-P_E m^DM_{R}^{-1}{m^D}^T P_E  =
V \mbox{diag}(0,e^{i\beta}m_2,m_3) V^{ T},
\eeq
or,
\beq
m^{D'}M_{R}^{-1}{m^{D'}}^T  =
V \mbox{diag}(0,e^{i\beta}m_2,m_3) V^{T},
\label{primed1}
\eeq
where,
\beq
\label{primed2}
m^{D'}=iP_Em^D=
i \left( \begin{array}{ccc}
e^{i\phi_{e}} & 0  & 0   \\
0 & e^{i\phi_{\mu}}  & 0 \\
0 & 0 & e^{i\phi_{\tau}}
\end{array}
\right) 
\left( \begin{array}{cc}
0 & a   \\
e & b \\
f & c 
\end{array}
\right).
\eeq
Finally, expanding Eq.\ref{primed1} using Eqs.~\ref{2rh1} and \ref{primed2},
we arrive at our ``master formula'',
\beq
\label{master}
\left( \begin{array}{ccc}
\tilde{a}^2 & \tilde{a}\tilde{b}   &  \tilde{a}\tilde{c}   \\
 \tilde{a}\tilde{b}  &  \tilde{e}^2+\tilde{b}^2  &  \tilde{e}\tilde{f}+ \tilde{b}\tilde{c}  \\
 \tilde{a}\tilde{c} & \tilde{e}\tilde{f}+ \tilde{b}\tilde{c} & \tilde{f}^2+\tilde{c}^2
\end{array}
\right)_{\alpha \beta} =
e^{i\beta}m_2V_{\alpha 2}V_{\beta 2}+
m_3V_{\alpha 3}V_{\beta 3},
\eeq
where we have defined
\beq
\label{primed4}
 \tilde{a} \equiv \frac{ie^{i\phi_{e}}a}{\sqrt{X}},  \ \  \tilde{b} \equiv \frac{ie^{i\phi_{\mu}}b}{\sqrt{X}},  
 \ \ \tilde{c} \equiv \frac{ie^{i\phi_{\tau}}c}{\sqrt{X}},
 \  \  \tilde{e} \equiv \frac{ie^{i\phi_{\mu}}e}{\sqrt{Y}}, \ \  \tilde{f} \equiv \frac{ie^{i\phi_{\tau}}f}{\sqrt{Y}}.
\eeq

The ``master formula'' in Eq.\ref{master} is a very useful equation, since it relates the re-phased Dirac mass
matrix parameters $a,b,c,e,f$, scaled by the positive square roots of the right-handed neutrino masses $X,Y$, 
which are the inputs of the see-saw mechanism (see Eq.~\ref{2rh1}), to physical
low energy observables. The combinations of see-saw input parameters appear on the LHS of Eq.\ref{master}
while the physical neutrino masses, mixing angles and phases appear on the RHS of Eq.\ref{master}.

In terms of parameter counting, we see that the RHS of the ``master formula''
Eq.\ref{master}, with $V$ in Eq.\ref{eq:matrix},
involves 7 non-zero physical parameters,
comprising 2 real positive masses $m_2,m_3$, the 3 mixing angles $\theta_{23}, \theta_{13}, \theta_{12}$, the oscillation phase $\delta$ and the single Majorana phase $\beta$, which determine exactly the 
5 complex parameters $ \tilde{a}, \tilde{b}, \tilde{c}, \tilde{e}, \tilde{f}$ on the LHS which therefore 
involve only two independent physical phases. 
We shall refer to $ \tilde{a}, \tilde{b}, \tilde{c}, \tilde{e}, \tilde{f}$ 
as the ``physical see-saw parameter combinations''.
Note that the 5 input complex Dirac masses $a,b,c,e,f$ also only depend on two physical phase differences which are
left invariant under charged lepton
re-phasing as discussed further later.

\subsection{Sequential Dominance and the Dominant Texture Zero}
So far we have considered the two right-handed neutrino model with a single texture zero in the electron
component, assuming a normal neutrino mass hierarchy, and have derived a 
``master formula'' in Eq.\ref{master}, relating see-saw parameters to physical observables.
As a first application of this formula, we show that SD is a consequence of 
the assumed texture zero $d=0$ which turns out to be the ``dominant texture zero''.

Writing out the ``master formula'' Eq.\ref{master} explicitly for all $(\alpha , \beta )$ gives,
\bea
\label{master2}
\tilde{a}^2 & =& e^{i\beta}m_2V_{1 2}^2+ m_3V_{1 3}^2,\\
\tilde{a}\tilde{b} & =& e^{i\beta}m_2V_{1 2}V_{2 2}+ m_3V_{1 3}V_{2 3},\\
\tilde{a}\tilde{c} & =& e^{i\beta}m_2V_{1 2}V_{3 2}+ m_3V_{1 3}V_{3 3},\\
\tilde{e}\tilde{f}+\tilde{b}\tilde{c}  & =& e^{i\beta}m_2V_{2 2}V_{3 2}+ m_3V_{2 3}V_{3 3},\\
\tilde{e}^2 + \tilde{b}^2 & =& e^{i\beta}m_2V_{2 2}^2+ m_3V_{2 3}^2,\\
\tilde{f}^2 + \tilde{c}^2 & =& e^{i\beta}m_2V_{3 2}^2+ m_3V_{3 3}^2. 
\label{master3}
\eea

We can solve for the individual parameters in terms of the above combinations,
\bea
\label{master3.4}
\tilde{b}^2 & =&  \frac{(\tilde{a}\tilde{b})^2}{\tilde{a}^2},\\
\tilde{c}^2 & =&  \frac{(\tilde{a}\tilde{c})^2}{\tilde{a}^2},\\
\tilde{e}^2 & =&  (\tilde{e}^2 + \tilde{b}^2)-\tilde{b}^2,\\
\tilde{f}^2 & =&  (\tilde{f}^2 + \tilde{c}^2)-\tilde{c}^2.
\label{master3.5}
\eea

Using Eqs.\ref{master2}-\ref{master3} and TBC mixing in Eq.\ref{TBC}, we estimate, to leading order in $\lambda$,
\bea
\label{master4}
\tilde{a}^2 & \sim & e^{i\beta} \frac{ m_2}{3},\\
\tilde{a}\tilde{b} & \sim &  e^{i\beta}\frac{ m_2}{3} + e^{-i\delta}\frac{ \lambda m_3}{2},\\
\tilde{a}\tilde{c} & \sim & -e^{i\beta}\frac{ m_2}{3} + e^{-i\delta}\frac{ \lambda m_3}{2},\\
\tilde{e}^2 + \tilde{b}^2 & \sim & \frac{ m_3}{2},\\
\tilde{f}^2 + \tilde{c}^2 & \sim & \frac{ m_3}{2}.
\label{master5}
\eea
Since the mass ratio may be approximated as $m_2/m_3 \approx \theta_{13}\approx  \lambda / \sqrt{2}$,
from Eqs.\ref{master4}-\ref{master5} we conclude that, 
\beq
(\tilde{e}+\tilde{f})^2\gg (\tilde{a}+\tilde{b}+\tilde{c})^2
\label{SD2}
\eeq
since the LHS is of order $m_3$, while the RHS is of order $m_2$. 
Using Eq.\ref{primed4}, we see that Eq.\ref{SD2} just corresponds to the SD condition in 
Eq.\ref{SDcond} where we identify the see-saw matrices in Eq.\ref{2rh1} with the notation
in Section~\ref{SD},
\begin{equation}
\left( \begin{array}{cc}
0 & a   \\
e & b \\
f & c 
\end{array}
\right)
\equiv
\left( \begin{array}{cc}
m^D_{e, \rm atm} & m^D_{e, \rm sol}   \\
m^D_{\mu, \rm atm} & m^D_{\mu, \rm sol}  \\
m^D_{\tau, \rm atm} & m^D_{\tau, \rm sol} 
\end{array}
\right),
\ \ \ \ 
\left( \begin{array}{cc}
Y & 0     \\
0 & X 
\end{array}
\right)
\equiv
\left( \begin{array}{cc}
M_{\rm atm} & 0    \\
0 & M_{\rm sol} 
\end{array}
\right),
\label{2rh3}
\end{equation}
from which we see that the texture zero in the electron component is in fact the 
``dominant texture zero'',
since we identify $m^D_{e, \rm atm}=0$ 
and $Y=M_{\rm atm}$. 
From our present perspective, where the reactor angle is measured,
we see that SD and the ``dominant texture zero'' implies and is implied by the 
data on the mixing angles which is approximated here by TBC mixing. 
However, historically, a decade before the measurement of $\theta_{13}$,
SD was simply postulated, along with the ``dominant texture zero''. This was then 
shown to lead to the bound on the reactor angle $\theta_{13} \simlt m_2/m_3$ 
using approximate analytic formulas for the 
mixing angles and phases expressed 
in terms of input see-saw parameters \cite{King:2002nf,King:2002qh}.


\subsection{Excluding a second texture zero  }
In this subsection we use the ``master formula'' in Eq.\ref{master}
to search for simple patterns of neutrino Dirac masses,
or equivalently neutrino Yukawa couplings, which could motivate some future flavour model,
for example based on a discrete family symmetry. 
To address such questions, it is relevant to take the ratios of the see-saw parameters
in the same Dirac column, so that the right-handed neutrino masses cancel.
From Eqs.\ref{master2}-\ref{master3} and Eq.\ref{primed4} we find,
\bea
\label{master6}
z_1 \equiv \frac{\tilde{e}}{\tilde{f}} & = &\pm
\frac{V_{23}V_{12}-V_{13}V_{22}}{V_{33}V_{12}-V_{13}V_{32}} \\
\label{master65}
z_2 \equiv \frac{\tilde{b}}{\tilde{a}} & = &
\frac{e^{i\beta}m_2V_{1 2}V_{2 2}+ m_3V_{1 3}V_{2 3}}
{e^{i\beta}m_2V_{1 2}^2+ m_3V_{1 3}^2}\\
z_3 \equiv \frac{\tilde{c}}{\tilde{a}}  &  = &
\frac{e^{i\beta}m_2V_{1 2}V_{3 2}+ m_3V_{1 3}V_{3 3}}
{e^{i\beta}m_2V_{1 2}^2+ m_3V_{1 3}^2}. 
\label{master66}
\eea
Note that $z_1$ is independent of neutrino masses (and phases).
It is also worth emphasising that, using Eq.\ref{primed4}, 
the moduli of these ratios are equal to the moduli of the original
Dirac mass matrix elements,
\beq
|z_1|=\frac{|e|}{|f|}, \ \ \ \ |z_2|=\frac{|b|}{|a|}, \ \ \ \ |z_3|=\frac{|c|}{|a|}. 
\label{ratios}
\eeq

For the case of TBC2 mixing in Eq.\ref{ansatzTBC2}, we find,
\bea
\label{V11}
V_{11} & = & \sqrt{\frac{2}{3}}(1+\frac{1}{4}\lambda^2)   \\
V_{12}& =&  \frac{1}{\sqrt{3}}(1-\frac{5}{4}\lambda^2) \\
V_{13}& =&  \frac{1}{\sqrt{2}}\lambda e^{-i\delta } \\
V_{21}& =&  -\frac{1}{\sqrt{6}}(1+\frac{1}{2}\lambda+\lambda e^{i\delta }-\frac{5}{4}\lambda^2- \frac{1}{2}\lambda^2 e^{i\delta })\\
V_{22}& =& \frac{1}{\sqrt{3}}(1+\frac{1}{2}\lambda- \frac{1}{2}\lambda e^{i\delta }+\frac{1}{4}\lambda^2+\frac{1}{4}\lambda^2e^{i\delta }) \\
V_{23}& =&  \frac{1}{\sqrt{2}}(1-\frac{1}{2}\lambda-\frac{1}{4}\lambda^2) \\
V_{31}& =& \frac{1}{\sqrt{6}}(1-\frac{1}{2}\lambda-\lambda e^{i\delta }-\lambda^2- \frac{1}{2}\lambda^2 e^{i\delta }) \\
V_{32}& =& -\frac{1}{\sqrt{3}}(1-\frac{1}{2}\lambda+\frac{1}{2}\lambda e^{i\delta }+\frac{1}{2}\lambda^2-\frac{1}{4}\lambda^2e^{i\delta })\\
V_{33}& =&  \frac{1}{\sqrt{2}}(1+\frac{1}{2}\lambda-\frac{1}{2}\lambda^2).
\label{V33}
\eea
We shall approximate the mass ratio as:
\beq
\frac{m_2}{m_3}\equiv \epsilon \approx  \frac{\lambda}{ \sqrt{2}} + \frac{1}{3}\lambda^2\approx 0.176.
\label{massratio}
\eeq
According to the global fits, which are compared in \cite{King:2013eh},
the mass ratio lies in the one sigma range $m_2/m_3=0.17-0.18$.

\begin{figure}[t]
\centering
\includegraphics[width=0.49\textwidth]{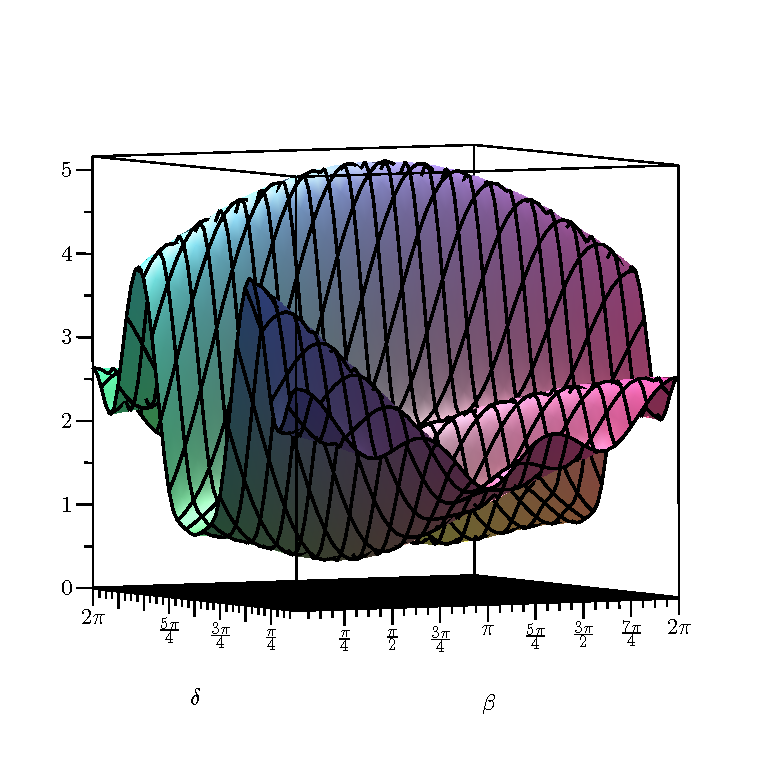}
\includegraphics[width=0.49\textwidth]{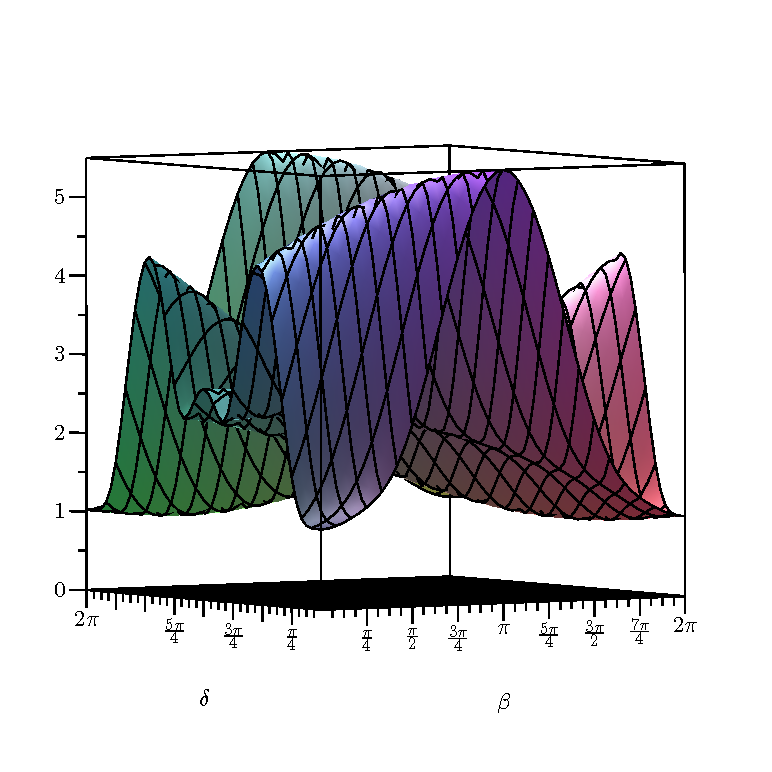}
\vspace*{-4mm}
    \caption{The left and right panels show $|z_2|$ and $|z_3|$ (respectively)
     as functions of the physical phases $\delta$ and $\beta$
     evaluated using Eqs.\ref{master65},\ref{master66} and using the TBC2 ansatz in 
     Eq.\ref{ansatz}, working to second order in $\lambda$.
     From these plots we see that a second texture zero corresponding to either $|z_2|=0$
     or $|z_3|=0$ is not possible.
     Similar results are obtained for TBC mixing.} \label{absz23}
\vspace*{-2mm}
\end{figure}

Using these approximations we perform an expansion of the ratios including terms of order $\lambda^2$,
which we use for our numerical results. However, since the full expansion is not very instructive, we only
write out the terms to order $\lambda$ below:
\bea
\label{master7}
z_1 \equiv \frac{\tilde{e}}{\tilde{f}} & = & \pm [1-\lambda (1+2e^{-i\delta })]+{\cal O}(\lambda^2)\\
z_2 \equiv \frac{\tilde{b}}{\tilde{a}}& = & 
 1+\frac{3}{\sqrt{2}}e^{-i(\beta+\delta ) }
+\frac{1}{2}\lambda (B-C)+{\cal O}(\lambda^2)  \label{master8}
 \\
z_3 \equiv \frac{\tilde{c}}{\tilde{a}}  & = &
 -1+\frac{3}{\sqrt{2}}e^{-i(\beta+\delta ) }
+\frac{1}{2}\lambda (B+C) +{\cal O}(\lambda^2),
\label{master9}
\eea
where we have introduced the complex notation,
\bea
B&=& 1 - e^{i\delta }
- 2e^{-i(\beta+\delta ) }-9e^{-i(2\beta+3\delta ) }\\
C&=&  3\sqrt{2}e^{-i(\beta+2\delta ) } +\frac{3}{\sqrt{2}}e^{-i(\beta+\delta ) } .
\eea

We can use these results to investigate the viability of a second texture zero.
Apparently a second texture zero with $|a|=0$ is inconsistent with Eqs.\ref{master8},\ref{master9}.
Also either $|e|=0$ or $|f|=0$  is inconsistent with Eq.\ref{master7}.
A second texture zero with either $|b|=0$ or $|c|=0$ would correspond to 
either $|z_2|=0$ or $|z_3|=0$.
This looks unlikely to be possible from the leading order results,
but requires numerical investigation to be sure.
We find that, for both TBC mixing and the more accurate TBC2
parameterisation given by Eqs.\ref{V11}-\ref{V33}, to second order in $\lambda$,
either $|z_2|=0$ or $|z_3|=0$ are not possible. This is illustrated for the second case of TBC2 mixing
in Fig.\ref{absz23}. In summary, we conclude that all the following two texture zero cases are not viable,
\bea
m^D &=&
\left( \begin{array}{cc}
0 & \times   \\
0  & \times  \\
\times  & \times  
\end{array}
\right), \ \ 
\left( \begin{array}{cc}
0  & \times  \\
\times  & \times  \\
0 & \times  
\end{array}
\right), \ \ 
\left( \begin{array}{cc}
0 & 0  \\
\times  & \times  \\
\times  & \times  
\end{array}
\right), \ \ 
\left( \begin{array}{cc}
0  & \times  \\
\times  & 0 \\
\times  & \times  
\end{array}
\right), 
 \ \ 
\left( \begin{array}{cc}
0  & \times  \\
\times  & \times  \\
\times  & 0
\end{array}
\right), \nonumber
\\
& &
\left( \begin{array}{cc}
\times & 0   \\
\times  & 0  \\
\times  & \times  
\end{array}
\right), \ \ 
\left( \begin{array}{cc}
\times  & 0   \\
\times  & \times  \\
\times  & 0 
\end{array}
\right),
 \ \ 
\left( \begin{array}{cc}
\times & 0  \\
0  & \times  \\
\times  & \times  
\end{array}
\right),
 \ \ 
\left( \begin{array}{cc}
\times & 0   \\
\times  & \times  \\
0  & \times  
\end{array}
\right). 
\eea

\section{Searching for simple ratios of Yukawa couplings and phases}
\label{search}


We shall write the neutrino 
Dirac mass matrix, or equivalently the neutrino Yukawa matrix, in some arbitrary phase basis 
(but in the diagonal right-handed neutrino and charged lepton mass basis) as,
\begin{equation}
\label{basis}
\left( \begin{array}{cc}
0 & |a|e^{i\phi_a}   \\
|e|e^{i\phi_e} & |b|e^{i\phi_b} \\
|f|e^{i\phi_f} & |c|e^{i\phi_c}  
\end{array}
\right),
\end{equation}
where the phase $\phi_a$ is not physical and can always be removed by charged lepton re-phasing,
while the two physical phases which are left invariant under charged lepton re-phasing are given,
using Eq.\ref{primed4}, by,
\beq
\eta_2=\phi_b-\phi_e=\arg{\tilde{b}}-\arg{\tilde{e}}, 
\ \ \ \ \eta_3= \phi_c-\phi_f=\arg{\tilde{c}}-\arg{\tilde{f}}.
\label{phibc}
\eeq
Note that the columns of Eq.\ref{basis} may be interchanged since the right-handed neutrino masses are arbitrary,
but the texture zero is always in the electron component of the ``dominant'' column as shown previously.
In order to determine the two physical phases $\eta_2$ and  $\eta_3$
we first need to determine the second order variables 
on the left hand side of Eqs.\ref{master2}-\ref{master3}.
From these we can then determine all the quadratic variables, 
$\tilde{a}^2$,$\tilde{b}^2$,$\tilde{c}^2$,$\tilde{e}^2$,$\tilde{f}^2$.
For example $\tilde{b}^2=(\tilde{a}\tilde{b})^2/\tilde{a}^2$.
There is a sign ambiguity in taking the square root of these quadratic variables
in order to find $\tilde{a}$,$\tilde{b}$,$\tilde{c}$,$\tilde{e}$,$\tilde{f}$ which we need to discuss.
We first note that the see-saw mechanism is invariant under a simultaneous
change of sign of $\tilde{a}, \tilde{b}, \tilde{c}$, or (independently) 
$\tilde{e}, \tilde{f}$, which corresponds to an independent undetermined sign
in the each of the two columns of the Dirac mass matrix.
Without loss of generality we shall use this sign freedom in the Dirac columns to fix $\tilde{b}$ and $\tilde{e}$
to be,
\beq
\arg{\tilde{b}}=\frac{1}{2}\arg{\tilde{b}^2}, 
\ \ \ \ \arg{\tilde{e}}=\frac{1}{2}\arg{\tilde{e}^2}, 
\label{phibe}
\eeq
so that these phases lie in the restricted range $-\pi/2$ to $\pi/2$,
rather than the full range $-\pi$ to $\pi$ over which $\arg{\tilde{b}^2}$ and $\arg{\tilde{e}^2}$ vary.
Then the phase of $\tilde{c}$ is determined using formulae
which include the information about relative phases of $\tilde{b}$ and $\tilde{c}$ in Eqs.\ref{master2}-\ref{master3},
\beq
\arg{\tilde{c}} = \frac{1}{2}\arg{\tilde{c}^2} - \arg\left( \frac{\tilde{a}\tilde{b}}{\tilde{a}\tilde{c}} \right),
\label{phic}
\eeq
(where the relative phase of $\tilde{b}$ and $\tilde{c}$ is meaningful since these variables are 
defined such that the charged lepton phases are fixed).
Having determined the phases of $\tilde{e}$, $\tilde{b}$ and $\tilde{c}$, we can then determine the phase
of $\tilde{f}$ using a formula which takes into account the relative phase of 
$\tilde{f}$ in the term involving $\tilde{e}\tilde{f}+\tilde{b}\tilde{c}$
in Eqs.\ref{master2}-\ref{master3}, which, after squaring and rearranging, yields the formula,
\beq
\arg{\tilde{f}} =-\arg{\tilde{e}}-\arg{\tilde{b}}-\arg{\tilde{c}}+\arg\frac{1}{2}\left[  
({\tilde{e}}{\tilde{f}}+{\tilde{b}}{\tilde{c}})^2 - {\tilde{e}}^2{\tilde{f}}^2 -  {\tilde{b}}^2{\tilde{c}}^2
\right].
\label{phif}
\eeq
Having thus correctly determined the phases of $\tilde{b}$,$\tilde{c}$,$\tilde{e}$,$\tilde{f}$,
then $\eta_2$, $\eta_3$ may be determined using Eq.\ref{phibc} together with
Eqs.\ref{phibe},\ref{phic},\ref{phif}.

\begin{figure}[h]
\centering
\includegraphics[width=0.49\textwidth]{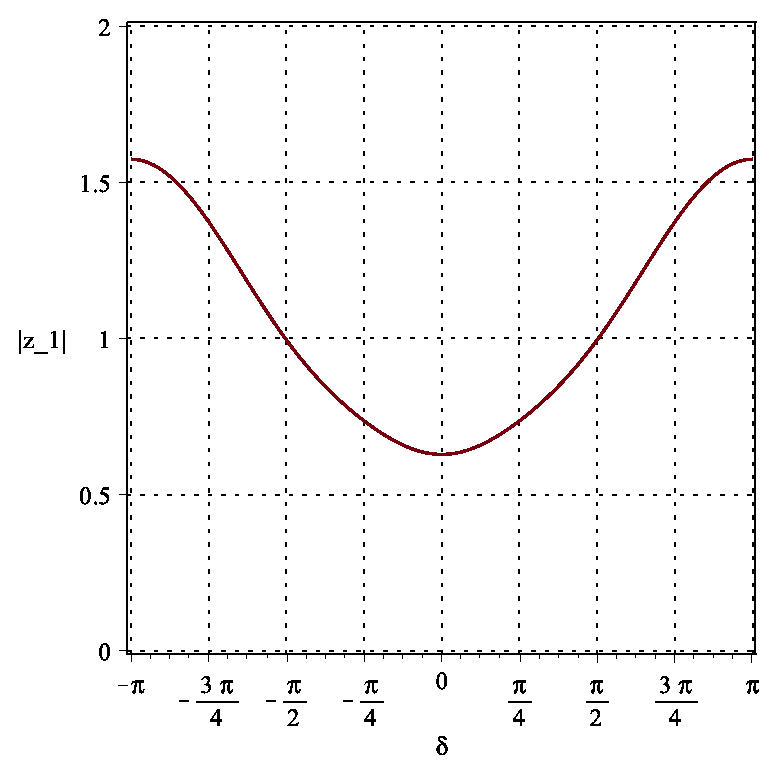}
\includegraphics[width=0.49\textwidth]{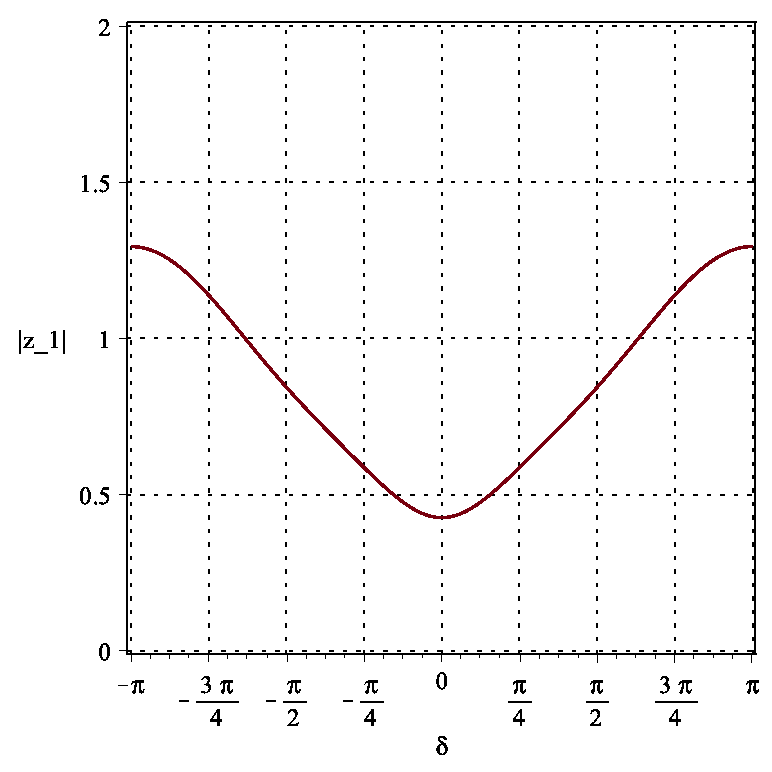}
\vspace*{-4mm}
    \caption{The curves show the ratio of Yukawa couplings in the dominant column $|z_1|=|e|/|f|$
     as functions of the physical phase $\delta$     
     evaluated using Eq.\ref{master6}.
     The left panel is for TBC mixing in Eq.\ref{TBC}
     while the right panel is for the TBC2 ansatz in Eq.\ref{ansatz}, both evaluated
     to second order in $\lambda$.
         From the left panel we see that $|e|=|f|$
     is possible for $\delta \approx \pm \pi/2$ for TBC mixing.
     From the right panel we see that $|e|/|f| \approx 1/2$ is possible for $\delta \approx 0$
    the TBC2 ansatz in Eq.\ref{ansatz}.} 
     \label{z1}
\vspace*{-2mm}
\end{figure}




Since we have excluded the possibility of a second texture zero in the previous subsection,
here we now consider other types of simple Yukawa coupling patterns, using the ``master formula'' 
in Eqs.\ref{master2}-\ref{master3}
combined with the simple (but accurate) parameterisations of mixing angles discussed above,
namely TBC in Eq.\ref{TBC} or TBC2 in Eq.\ref{ansatz}.
We first recall the meaning of the ratios $|z_i|$ in Eq.\ref{ratios}, namely that 
they correspond to the ratios of magnitudes of Dirac mass matrix elements, or equivalently, Yukawa couplings,
and we repeat this result for convenience,
\beq
|z_1|=\frac{|e|}{|f|}, \ \ \ \ |z_2|=\frac{|b|}{|a|}, \ \ \ \ |z_3|=\frac{|c|}{|a|}. 
\label{ratios2}
\eeq
These ratios will be calculated in this subsection using Eqs.\ref{master6}-\ref{master66}.

Recall that the ratios of Yukawa couplings in the dominant column,
$z_1$ in Eq.\ref{master6}, is independent of neutrino masses (and phases) and it only depends on the mixing angles and oscillation phase.
In Fig.\ref{z1} we see that $|z_1|=1$ or $|z_1|=1/2$ are possible for particular choices of physical phase $\delta$ which depends on the precise values of the mixing angles.
From the left panel of Fig.\ref{z1} 
we see that for TBC mixing we can achieve
\beq
|z_1|=\frac{|e|}{|f|}=1 
\label{z1equals1}
\eeq
for a particular choice of oscillation phase,
$\delta = \pm \pi /2$. From the right panel of Fig.\ref{z1} we see that for TBC2 mixing we can achieve 
\beq
|z_1|=\frac{|e|}{|f|}=\frac{1}{2}
\label{z1equalshalf}
\eeq
for a particular choice of oscillation phase close to zero, $\delta \approx 0$.

\begin{figure}[p]
\centering
\includegraphics[width=0.49\textwidth]{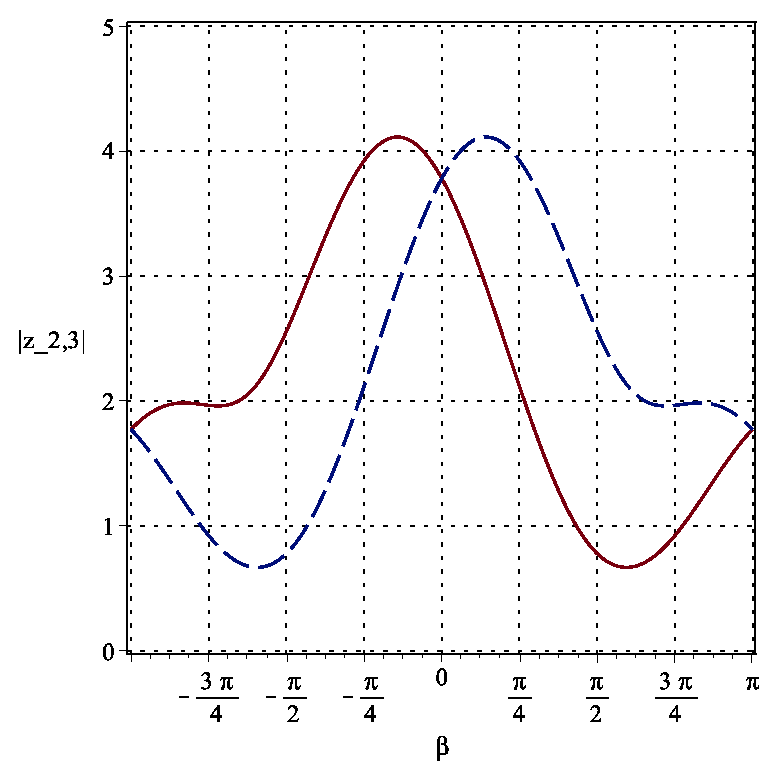}
\includegraphics[width=0.49\textwidth]{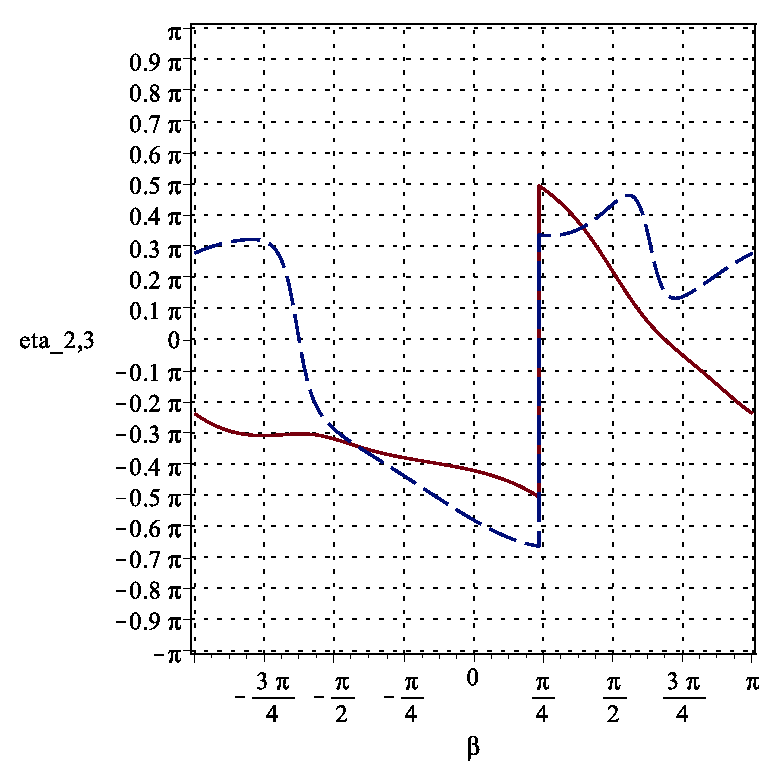}
\vspace*{-4mm}
    \caption{Results for TBC mixing in Eq.\ref{TBC} 
    with $\delta =\pi/2$ and $|z_1|=|e|/|f|=1$ as functions of the physical Majorana phase $\beta$.
    The left panel shows the ratios of Dirac mass matrix elements $|z_2|=|b|/|a|$ (solid) and $|z_3|=|c|/|a|$ (dashes).
     The right panel shows the physical phase differences $\eta_2=\phi_b-\phi_e$ (solid) and $\eta_3= \phi_c-\phi_f$ (dashes). Related results may be obtained for $\delta =-\pi/2$ and  $|z_1|=|e|/|f|=1$.
             } \label{TBCdeltapiover2}
\vspace*{-2mm}
\end{figure}

\begin{table}[p]
	\centering
		\begin{tabular}{|c||c|c||c|c|c|c|c|}
			\hline
			Mixing & $\delta$ &  $|z_1|$ & $\beta$  &$|z_2|$ & $|z_3|$ & $\eta_2$ & $\eta_3$ \\ \hline \hline
			TBC & $\pm \pi/2$ & 1& 0 & 4 & 4 & $\mp 2\pi/5$ & $\mp 3\pi/5$ \\ \hline
			TBC & $\pm \pi/2$ & 1& $\pm \pi/4$ & 2 & 4 & $\pm \pi/2$ & $\pm \pi/3$ \\ \hline
			TBC & $\pm \pi/2$ & 1& $\mp \pi/4$ & 4 & 2 & $\mp 2\pi/5$ & $\mp 2\pi/5$ \\ \hline
			TBC & $\pm \pi/2$ & 1& $\pm 3\pi/8$ & 1 & 3 & $\pm \pi/3$ & $\pm \pi/3$ \\ \hline
			TBC & $\pm \pi/2$ & 1& $\mp 3\pi/8$ & 3 & 1 & $\mp \pi/3$ & $\mp \pi/3$ \\ \hline
			TBC & $\pm \pi/2$ & 1& $\pm \pi/2$ & 1 & 2.5 & $\pm \pi/5$ & $\pm 2\pi/5$ \\ \hline
			TBC & $\pm \pi/2$ & 1& $\mp \pi/2$ & 2.5 & 1 & $\mp \pi/3$ & $\mp \pi/3$ \\ \hline	
			TBC & $\pm \pi/2$ & 1& $\pm 3\pi/4$ & 1 & 2 & 0 & $\pm \pi/6$ \\ \hline
			TBC & $\pm \pi/2$ & 1& $\mp 3\pi/4$ & 2 & 1 & $\mp \pi/3$ & $\pm \pi/3$ \\ \hline
			TBC & $\pm \pi/2$ & 1& $\pm \pi$ & 2 & 2 &$\mp \pi/5$ & $\pm \pi/3$ \\ \hline
                         TBC & $\pm \pi/2$ & 1& $\mp \pi$ & 2 & 2 & $\mp \pi/5$ & $\pm \pi/3$ \\ \hline                    
				\end{tabular} 
			\caption{Examples of simple ratios of Yukawa couplings 
			$|z_2|=|b|/|a|$, $|z_3|=|c|/|a|$
			and corresponding approximate phases
			as estimated from Fig.\ref{TBCdeltapiover2}.
			The exact numerical PMNS parameters for selected cases will be considered later.
			}
		\label{TBCtable}
\end{table}

\begin{figure}[p]
\centering
\includegraphics[width=0.49\textwidth]{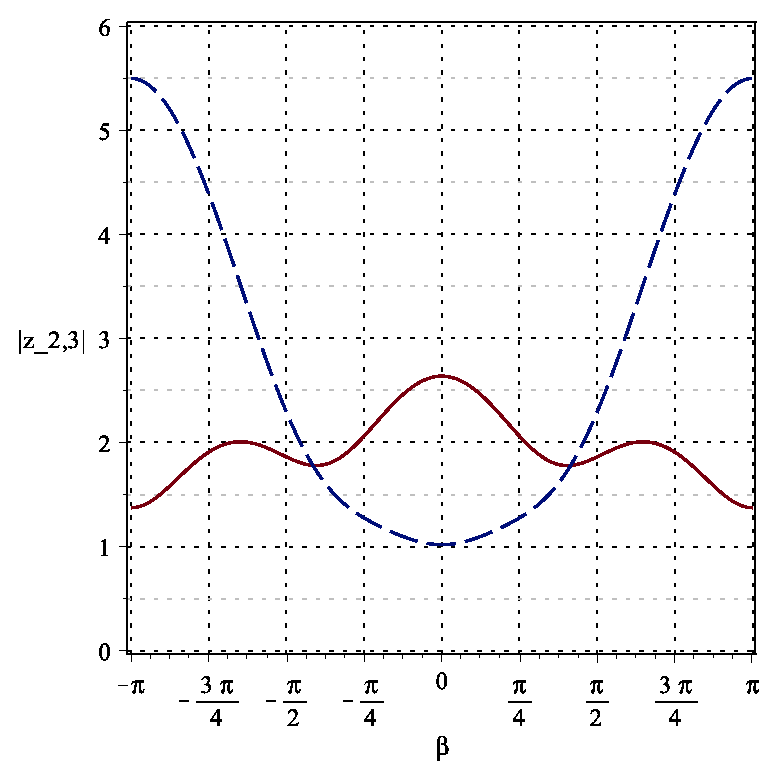}
\includegraphics[width=0.49\textwidth]{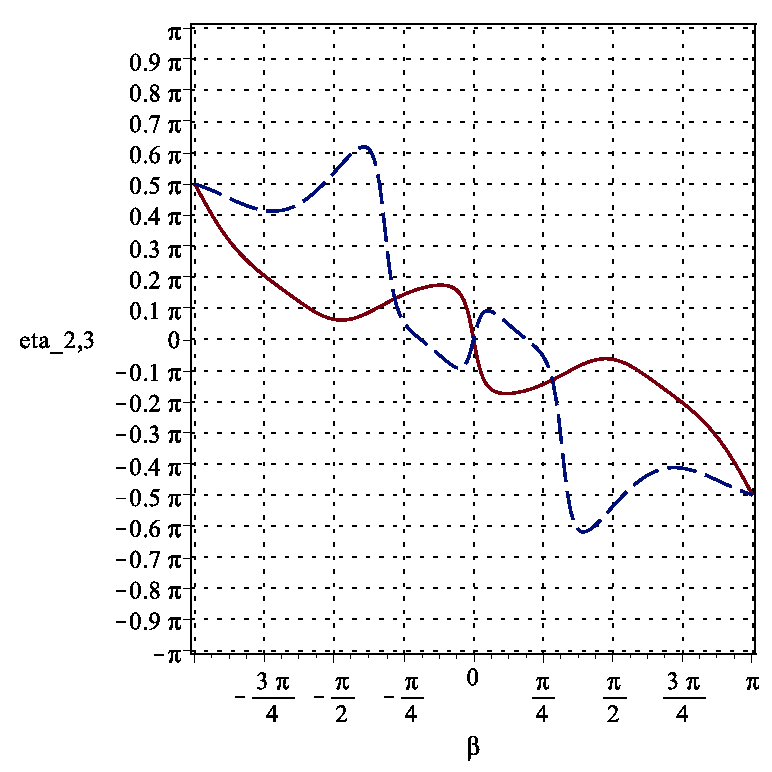}
\vspace*{-4mm}
    \caption{Results for TBC2 mixing  in Eq.\ref{ansatz}
    with $\delta =0$ and $|z_1|=|e|/|f|=1/2$ as functions of the physical Majorana phase $\beta$.
    The left panel shows the ratios of Dirac mass matrix elements $|z_2|=|b|/|a|$ (solid) and $|z_3|=|c|/|a|$ (dashes).
     The right panel shows the physical phase differences $\eta_2=\phi_b-\phi_e$ (solid) and $\eta_3= \phi_c-\phi_f$ (dashes).
             } \label{TBC2deltazero}
\vspace*{-2mm}
\end{figure}

\begin{table}[p]
	\centering
		\begin{tabular}{|c||c|c||c|c|c|c|c|}
			\hline
			Mixing & $\delta$ &  $|z_1|$ & $\beta$  &$|z_2|$ & $|z_3|$ & $\eta_2$ & $\eta_3$ \\ \hline \hline
			 TBC2 & $0$ & 0.5 & $0$ & 2.5 & 1 & $0$ & $0$ \\ \hline       
			 TBC2 & $0$ & 0.5 & $\pm \pi/4$ & 2 & 1 & $\mp \pi/6$ & $0$ \\ \hline  
			 TBC2 & $0$ & 0.5 & $\pm \pi/2$ & 2 & 2 & $0$ & $\mp \pi/2$ \\ \hline   
			 TBC2 & $0$ & 0.5 & $\pm 3\pi/4$ & 2 & 4.5 & $\mp \pi/5$ & $\mp 2\pi/5$ \\ \hline    
			  TBC2 & $0$ & 0.5 & $\pm \pi$ & 1.5 & 5.5 & $\mp \pi/2$ & $\mp \pi/2$ \\ \hline          
				\end{tabular} 
			\caption{Examples of simple ratios of Yukawa couplings $|z_2|=|b|/|a|$, $|z_3|=|c|/|a|$
			and corresponding approximate phases as estimated from Fig.\ref{TBC2deltazero}.
			The exact numerical PMNS parameters for selected cases will be considered later.
			}
		\label{TBC2table}
\end{table}

These are quite interesting results, namely that simple ratios such as $|e|=|f|$ or $|e|/|f|=1/2$
are associated with simple values of oscillation phase, so we shall focus on these two examples.

In the left panel of Fig.\ref{TBCdeltapiover2} we show the absolute magnitudes of the ratios $|z_2|=|b|/|a|$ and $|z_3|=|c|/|a|$ as a function of the Majorana phase $\beta$,
for the case of TBC mixing, keeping the oscillation phase fixed at $\delta =\pi /2$ corresponding to $|e|=|f|$
in Eq.\ref{z1equals1}.
These ratios concern the second subdominant column of the Dirac mass matrix.
As well as simple values of these ratios we are also interested in simple values of the physical phase differences $\eta_2=\phi_b-\phi_e$ and $\eta_3= \phi_c-\phi_f$ and these are shown in the right panel of Fig.\ref{TBCdeltapiover2}as a function of the Majorana phase $\beta$. 

By examining the results in Fig.\ref{TBCdeltapiover2} we find some examples of simple ratios of Yukawa couplings in the subdominant column, which are shown together with the phases
in Table~\ref{TBCtable}. Analogous results for $\delta =-\pi /2$ are also shown in Table~\ref{TBCtable}, where these results correspond to a symmetry under a change of the sign of all the phases $\delta$, $\beta$, $\eta_2$, $\eta_3$.
This symmetry of the results can be understood by taking the complex conjugate of both sides of the ``master formula'' in Eq.\ref{master}. We have verified this symmetry numerically.

Similarly, in the left panel of Fig.\ref{TBC2deltazero} we show the absolute magnitudes of the ratios $|z_2|=|b|/|a|$ and $|z_3|=|c|/|a|$ as a function of the Majorana phase $\beta$,
for the case of TBC2 mixing, keeping the oscillation phase fixed at $\delta =0$ corresponding to $|e|/|f|=1/2$
in Eq.\ref{z1equalshalf}.
Simple values of the physical phase differences $\eta_2=\phi_b-\phi_e$ and $\eta_3= \phi_c-\phi_f$ are shown in the right panel of Fig.\ref{TBC2deltazero} as a function of the Majorana phase $\beta$. 
By examining the results in Fig.\ref{TBC2deltazero} we find some examples of simple ratios of Yukawa couplings in the subdominant column, which are shown together with the phases
in Table~\ref{TBC2table}. 

\section{Brief Review of Indirect Models}
\label{models}
The results in the previous section can be used as the starting point to explore new types of indirect models.
Indirect models are fully reviewed in \cite{King:2013eh} so here we only provide a brief review for completeness.

Indirect models start from simple forms of the Yukawa columns which can be obtained from vacuum alignment due to some discrete non-Abelian family symmetry which is however completely broken. 
In the class of indirect models, the Klein symmetry of the neutrino mass matrix is not identified as 
a subgroup of the non-Abelian discrete family symmetry~$G$. Models of this class are
typically based on the type~I seesaw mechanism together with
sequential dominance. Here, the main role
of the family symmetry consists in relating the Yukawa couplings $e,f$
as well as $a,b,c$ of Eq.~\ref{2rh3} by introducing 
flavon fields which transform as triplets under~$G$ (or anti-triplets if the representation is complex) and 
acquire special vacuum configurations. The directions of
the flavon alignments are determined by the $G$ symmetric operators of the
flavon potential~\cite{King:2013eh}.

Continuing to work in the basis where both the charged leptons as well as the right-handed
neutrinos are diagonal, the leptonic flavour structure is encoded in the Dirac
neutrino Yukawa operator. The triplet (or anti-triplet) flavons in the neutrino sector denoted 
$\phi_{\rm atm}$,  $\phi_{\rm sol}$ enter
linearly as,
\beq
{\cal L}^{\nu}\sim 
\frac{\phi_{\rm atm}}{\Lambda}\overline{L} N_{\rm atm} H + 
\frac{\phi_{\rm sol}}{\Lambda}\overline{L} N_{\rm sol} H
+ M_{\rm atm}\overline{N^c}_{\rm atm}N_{\rm atm}
+ M_{\rm sol}\overline{N^c}_{\rm sol}N_{\rm sol}
+ {H.c}
\label{indirect1}
\eeq
where $\Lambda$ is a cut-off scale and $H$ is the Standard Model Higgs doublet which is a singlet of $G$,
while $L$ are the three lepton doublets which form a triplet representation of $G$. 
The right-handed neutrinos $N_{\rm atm},N_{\rm sol}$ and the Higgs doublet $H$
are assumed to be in the singlet representation of $G$.
However to obtain only the above terms will require additional ``shaping symmetries'' which we do not specify
here ( for examples which also include the charged lepton sector see the review~\cite{King:2013eh}).
We obtain the Dirac neutrino
Yukawa matrix by inserting the flavon VEVs into
Eq.~\ref{indirect1}. Suppressing the dimensionless couplings of the Dirac
neutrino terms for notational clarity, we get 
\be
Y^\nu v ~=~ m_D  ~=~
\begin{pmatrix} 
d&a\\
e&b\\
f&c
\end{pmatrix}
\equiv
\left( \begin{array}{cc}
m^D_{e, \rm atm} & m^D_{e, \rm sol}   \\
m^D_{\mu, \rm atm} & m^D_{\mu, \rm sol}  \\
m^D_{\tau, \rm atm} & m^D_{\tau, \rm sol} 
\end{array}
\right)
\sim
\frac{v}{\Lambda}
\begin{pmatrix}
\vev{\phi_{\rm atm}}_1^{}& 
\vev{\phi_{\rm sol}}_1^{}\\
\vev{\phi_{\rm atm}}_2^{}& 
\vev{\phi_{\rm sol}}_2^{}\\
\vev{\phi_{\rm atm}}_3^{}& 
\vev{\phi_{\rm sol}}_3^{}
\end{pmatrix}.\label{indirect2}
\ee
The columns of the Dirac neutrino Yukawa matrix are therefore proportional to
the vacuum alignments of the flavons fields $\phi_{\rm atm}$,  $\phi_{\rm sol}$. The effective
Majorana operators of the light neutrinos can be derived from 
this using the seesaw formula of Eq.~(\ref{seesaw}), yielding
\be
{\cal L}^{\nu}_{eff}~\sim~  \overline{L} 
\left(
\frac{  \vev{\phi_{\rm atm}}   }{\Lambda} \cdot 
\frac{1}{M_{\rm atm}}  \cdot
\frac{  \vev{\phi_{\rm atm}}^T   }{\Lambda}
+
\frac{  \vev{\phi_{\rm sol}}   }{\Lambda} \cdot 
\frac{1}{M_{\rm sol}}  \cdot
\frac{  \vev{\phi_{\rm sol}}^T   }{\Lambda}
\right)
L^c  HH,  
\label{eq:indirectSQ}
\ee
which has the structure of Eq.\ref{leff} after the Higgs vev $v$ is inserted.
Note that the flavons enter the effective neutrino mass terms quadratically.

In the class of indirect models, the PMNS mixing pattern thus
becomes a question of the alignment vectors $\vev{\phi_{\rm atm}}$ and $\vev{\phi_{\rm sol}}$. 
For instance,
a neutrino mass matrix that gives rise to tri-bimaximal mixing can be
obtained using the flavon alignments of constrained sequential dominance (CSD) \cite{King:2005bj},
\begin{equation}
\label{Phi0} 
\frac{\vev{\phi_{\rm atm}}}{\Lambda}
=\frac{1}{\Lambda}\begin{pmatrix}
\vev{\phi_{\rm atm}}_1^{}\\
\vev{\phi_{\rm atm}}_2^{}\\
\vev{\phi_{\rm atm}}_3^{}
\end{pmatrix}
\propto \begin{pmatrix}0 \\ 1 \\ 1\end{pmatrix}e^{i\phi_e},
 \qquad
\frac{\vev{\phi_{\rm sol}}}{\Lambda}
=\frac{1}{\Lambda}\begin{pmatrix}
\vev{\phi_{\rm sol}}_1^{}\\
\vev{\phi_{\rm sol}}_2^{}\\
\vev{\phi_{\rm sol}}_3^{}
\end{pmatrix}
\propto \begin{pmatrix}1 \\ 1 \\ -1\end{pmatrix}e^{i\phi_b}.
\end{equation}
Note that the resulting columns of the Dirac neutrino Yukawa matrix 
in Eq.\ref{indirect2} are
proportional to the columns of the unitary (in the present case TB )
mixing matrix with the physical phase difference $\phi_b-\phi_e$
being related to the Majorana phase $\beta$. 
Such a property of the Dirac neutrino Yukawa matrix is
generally called form dominance~\cite{Chen:2009um}.\footnote{Exact form
dominance implies vanishing leptogenesis~\cite{Choubey:2010vs}.
This provides an independent motivation for deviating from CSD.} 
Furthermore these alignments are left invariant under the action of the
Klein symmetry generators, up to an irrelevant sign which
drops out due to the quadratic appearance of each flavon. 
Since the family symmetry $G$ does not contain the
neutrino Klein symmetry, its primary role is then to explain the origin of
these or similarly simple flavon alignments.  

It is clear that CSD is not realistic, primarily because it 
predicts TB mixing which involves a zero reactor angle.
However the above example of CSD does illustrate the general strategy of obtaining
simple ratios of Yukawa couplings within a particular column, including the 
texture zero $d=0$, from vacuum alignment within the framework of indirect models
using the techniques reviewed in~\cite{King:2013eh}.

An alternative approach to switching on the reactor angle without any ``texture zeros''
is to consider vacuum alignments 
which preserve the structure of CSD except that they allow $d\neq 0$, where such models 
are referred to as partially constrained sequential dominance (PCSD) \cite{King:2009qt}.
The required vacuum alignment has been discussed within an explicit $A_4$ model in \cite{King:2011ab}.
With the choice,
\begin{equation}
\label{Phi1} 
\frac{\vev{\phi_{\rm atm}}}{\Lambda}
\propto  \begin{pmatrix}re^{i\delta_0} \\ 1 \\ 1\end{pmatrix}e^{i\phi_e},
 \qquad
\frac{\vev{\phi_{\rm sol}}}{\Lambda}
\propto \begin{pmatrix}1 \\ 1 \\ -1\end{pmatrix}e^{i\phi_b},
\end{equation}
one obtains tri-bimaximal-reactor (TBR) mixing,
which corresponds to TB mixing but with a non-zero reactor deviation parameter $r$ 
and an oscillation phase $\delta \approx \delta_0$ \cite{King:2009qt,King:2011ab}.
For the choice $r=\lambda$ one arrives at TBC mixing~\cite{King:2012vj}.
However the main challenge for this approach is to explain why $r=\lambda$, 
which is not easy to do (for an attempt see~\cite{King:2012vj}).

Another alternative approach is to consider particular alignments with two  ``texture zeros''.
An example of this kind is,
\begin{equation}
\label{Phi2} 
\frac{\vev{\phi_{\rm atm}}}{\Lambda}
\propto  \begin{pmatrix}0 \\ 1 \\ 1\end{pmatrix}e^{i\phi_e},
 \qquad
\frac{\vev{\phi_{\rm sol}}}{\Lambda}
\propto  
 \begin{pmatrix}1 \\ 2 \\ 0\end{pmatrix}e^{i\phi_b}
 \ \ 
 {\rm or}
\ \ 
\begin{pmatrix}1 \\ 0 \\ -2\end{pmatrix}e^{i\phi_b},
\end{equation}
These two alignment possibilities which involve two ``texture zeros''
were called CSD2 \cite{Antusch:2011ic}. 
Note that the above alignments involve the components of each of the flavon alignments 
$\vev{\phi_{\rm atm}}$ and $\vev{\phi_{\rm sol}}$ being relatively real.
However the overall phase difference between the flavon alignments $\phi_b-\phi_e$ is physically
significant and particular values of this phase difference corresponding to multiples of $\pi/4$
are preferred by the numerical fit resulting in maximal leptonic CP violation \cite{Antusch:2011ic}.
Unfortunately, as we have seen, it is not possible to obtain a large
enough reactor angle for the case of a normal neutrino mass hierarchy with two texture zeros. Indeed the
reactor angle predicted from CSD2 was $\theta_{13}\approx \frac{\sqrt{2}}{3}\frac{m_2}{m_3}$
leading to $\theta_{13}\sim 5^o-6^o$ which is too small \cite{Antusch:2011ic}. The situation may be improved
by considering charged lepton corrections \cite{Antusch:2013wn}, but here we are considering the simple case where such corrections are completely negligible and in this case we cannot have two texture zeros in the two right-handed neutrino case of a normal neutrino mass hierarchy.

Our approach here is to maintain a single ``texture zero'' $d=0$ which leads to a qualitative understanding of the 
reactor angle, $\theta_{13} \simlt m_2/m_3$. However in order to provide a quantitative understanding of the 
lepton mixing angles, in the next section we shall consider 
particular types of vacuum alignment with $\vev{\phi_{\rm atm}}_1^{}=0$.

\section{Searching for New Types of Indirect Models}
\label{models2}
In the previous subsection we briefly discussed the approaches which already exist in the literature,
namely CSD, CSD2 and PCSD, which form the starting point of indirect models. In this section we shall use the results in this paper to explore new approaches which could form the starting point for new indirect models.
The new models will involve one single dominant texture zero, as in the case of CSD, and are unlike CSD2 or PCSD
which either involve two texture zeros or no texture zeros.
However, whereas CSD predicts a zero reactor angle, the new approaches will predict a non-zero reactor angle
consistent with experiment.

To obtain realistic solutions which can accommodate the non-zero reactor angle, while maintaining
the dominant texture zero $d=\vev{\phi_{\rm atm}}_1^{}=0$, we shall turn to the
results of the previous section based on TBC and TBC2 mixing which are representative of the results from
the latest global fits of leptonic mixing angles for the case of a normal neutrino mass hierarchy.
From the point of view of Indirect Models we are looking for new examples where the 
magnitudes of the Yukawa couplings in a particular column occur in simple ratios.

We emphasise that we only use TBC and TBC2 mixing to guide the search for simple alignments since they correspond to two phenomenologically acceptable mixing patterns. We thus allow simple alignments which 
only approximately satisfy TBC and TBC2 mixing, and may be taken as the starting point for new models,
leading to predictions for lepton mixing which do not fit exactly into either of these mixing patterns, but which we expect to be phenomenologically acceptable.

We shall work in a phase basis where the dominant column involves Yukawa couplings which are real and positive, so all the phases occur in the second subdominant column. With this phase convention we can write,
the Dirac mass (or Yukawa) matrix, previously given in Eq.\ref{basis}, together with Eq.\ref{indirect2},
as,
\begin{equation}
\label{basis2}
\left( \begin{array}{cc}
0 & |a|e^{i\phi_a}   \\
|e|& |b|e^{i\phi_b} \\
|f| & |c|e^{i\phi_c}  
\end{array}
\right)\equiv
\frac{v}{\Lambda}
\begin{pmatrix}
0 & 
\vev{\phi_{\rm sol}}_1^{}\\
\vev{\phi_{\rm atm}}_2^{}& 
\vev{\phi_{\rm sol}}_2^{}\\
\vev{\phi_{\rm atm}}_3^{}& 
\vev{\phi_{\rm sol}}_3^{}
\end{pmatrix},
\end{equation}
where the phase $\phi_a$ is not physical and can always be adjusted to take any value by charged lepton re-phasing,
while the two physical phases which are left invariant under charged lepton re-phasing,
previously given in Eq.\ref{phibc}, in this convention become,
\beq
\eta_2=\phi_b, 
\ \ \ \ \eta_3= \phi_c.
\label{phibc2}
\eeq
In this convention the flavon alignment $\vev{\phi_{\rm atm}}$ is purely real
with a zero value in its first component and no phase difference between its second and third
components, while the flavon alignment vector $\vev{\phi_{\rm sol}}$ may 
in principle involve three phases, one for each component.
Vacuum alignment techniques may lead to an arbitrary phase 
difference between different aligned components of the same flavon,
as in the case of PCSD discussed in Eq.\ref{Phi1} where the phase $\delta_0$ is undetermined.
However it is interesting to look for examples where
$\eta_2$ and $\eta_3$ are are simply related, differing only by a simple rational multiple of $\pi$,
since such simple relationships may be later understood in the context of some future theory.
One particularly simple example would be where the components 
$\vev{\phi_{\rm sol}}_1^{}$, $\vev{\phi_{\rm sol}}_2^{}$, $\vev{\phi_{\rm sol}}_3^{}$ are relatively real,
i.e. have the same phases up to $\pm \pi$ (remembering that the phase of $\vev{\phi_{\rm sol}}_1^{}$ is a free choice). An example of this kind of vacuum alignment is CSD2 in Eq.\ref{Phi2}.
A particularly simple example of this kind is,
\beq
\eta = \eta_2 = \eta_3 .
\label{eta0}
\eeq

In terms of general flavon models, 
the overall phases of the different flavon vevs $\vev{\phi_{\rm atm}}$ and $\vev{\phi_{\rm sol}}$
(i.e. the phases which are factored out of the alignments, for example
the remaining phases which appear in CSD in Eq.\ref{Phi0} or in CSD2 in Eq.\ref{Phi2})
are not determined since these overall phases depend on both arbitrary complex
Yukawa coupling constants and complex flavon vevs. This is unfortunate since physical mixing 
angles and CP violating phases depend on this difference of overall phases,
as in the example of CSD2 discussed above.
However, within special classes of models 
based on spontaneous CP violation, the Yukawa couplings are constrained to be real and in addition
the phases of the overall flavon vevs may be determined using the vacuum alignment techniques discussed in 
\cite{Antusch:2011sx}. Within such classes of models 
the overall phases of $\vev{\phi_{\rm atm}}$ and $\vev{\phi_{\rm sol}}$
may be determined related in terms of simple rational multiples
of $\pi$. Therefore we shall also be interested in such cases.

From Table~\ref{TBCtable} we find the following phenomenological possibility,
corresponding to TBC mixing with $\delta =\pi/2$ and $\beta =-\pi/4$,
where this case corresponds to $\eta = \eta_2=\eta_3=-2\pi/5$, 
$|z_1|=1$, $|z_2|=4$, $|z_3|=2$.
Using Eqs.\ref{basis2} and \ref{phibc2} we can see that 
this phenomenological example may be used as the basis for constructing a flavon model 
where the relative phases of the flavons are real, and the overall phases of the flavons take
simple values, namely zero and $-2i\pi/5$,
\begin{equation}
\label{Phi3} 
\frac{\vev{\phi_{\rm atm}}}{\Lambda}
\propto \begin{pmatrix} 0 \\ 1 \\ 1\end{pmatrix}\equiv A,
 \qquad
\frac{\vev{\phi_{\rm sol}}}{\Lambda}
\propto  \begin{pmatrix}1 \\ 4 \\ 2\end{pmatrix}e^{-2i\pi/5}\equiv B,
\end{equation}
where we have used the phase freedom in $\phi_a$ to set this phase equal to $-2\pi/5$.

The alignments defined as $A$ and $B$ above are not sufficient to determine
the effective neutrino mass matrix and hence the PMNS mixing parameters. 
Their normalisation is also required.
In the case of form dominance, the alignments would be sufficient to determine the PMNS mixing parameters
but form dominance is not satisfied.
The effective neutrino mass matrix can be written as,
\beq
m^{\nu}=m_aA^TA+m_bB^TB=m_aA^TA+m_a\epsilon_{\nu}B^TB
\label{seesaw2}
\eeq
where $m_a$ and $m_b$ are real mass parameters which determine the physical neutrino masses 
$m_3$ and $m_2$. For fixed alignments $A,B$, the ratio $\epsilon_{\nu}=m_b/m_a$ will also affect
the PMNS mixing parameters. For the above alignment this ratio is given by,
\beq
\epsilon_{\nu}=\left(\frac{1}{4}\frac{|\tilde{b}|}{|\tilde{e}|}\right)^2,
\eeq
where this ratio is related to physical parameters using Eqs.\ref{master3.4}-\ref{master3.5}.
For the case of TBC mixing with $\delta =\pi/2$ and $\beta =-\pi/4$, with
the mass ratio as in Eq.\ref{massratio}, we find numerically that 
\beq
\frac{|\tilde{b}|}{|\tilde{e}|}=0.9159.
\eeq
This completely fixes the neutrino mass matrix up to an overall scale $m_a$ which does not affect the
PMNS mixing parameters. Using the Mixing Parameter Tools (MPT) package associated 
with \cite{Antusch:2005gp}
\footnote{Note that the convention in \cite{Antusch:2005gp} for the neutrino mass
matrix $m^{\nu}$ differ from ours in Eq.\ref{Leff} by an overall complex conjugation. Also the convention
in \cite{Antusch:2005gp} for the Majorana phases differs from ours in Eq.\ref{MNS} by a
further complex conjugation.}
, we can check the predictions of this vacuum alignment,
which may be compared to the original mixing pattern that was assumed to derive it (in brackets).
For example, using the above alignment with $\epsilon_{\nu}=0.058$
we find using MPT, $m_2/m_3=0.17$ together with,
\beq
\theta_{12}=34.2^o , \ \ \theta_{13}=9.2^o ,\ \ \theta_{23}=40.9^o,  \ \ \delta = 107^o, \ \ \beta = -83^o,
\label{values1}
\eeq
which is in good agreement with the global fits. 
The discrepancy between Eqs.\ref{values1} and the TBC mixing angles in Eq.\ref{TBCangles}
arises due to the approximation in setting the ratios $|z_i|$ and the phases $\eta_i$
equal to simple values. 
We have checked that using the accurate values of the alignments, obtained from the master formula
in Eq.\ref{master},
with the ratios of Yukawa couplings calculated using Eqs.\ref{master6}, \ref{master65}, \ref{master7}
and the phases calculated using Eq.\ref{phibc} together with
Eqs.\ref{phibe},\ref{phic},\ref{phif}, namely:
\begin{equation}
\label{Phi4} 
A \equiv  \begin{pmatrix} 0 \\ 1 \\ 1\end{pmatrix},
 \qquad
 B \equiv  \begin{pmatrix}1 \\ 3.725 e^{-i\pi 0.373} \\ 2.028 e^{-i\pi 0.479} \end{pmatrix},
\end{equation}
with 
\beq
\epsilon_{\nu}=\left(\frac{1}{3.725}\frac{|\tilde{b}|}{|\tilde{e}|}\right)^2=0.06045,
\eeq
when used in Eq.\ref{seesaw2} reproduce the TBC angles in Eq.\ref{TBCangles}
to excellent accuracy.
This gives us confidence that the master formula and the procedure that we followed to obtain the
Yukawa ratios and the phases is correct. It also supports our approach of using the values obtained in 
Table~\ref{TBCtable} and Table~\ref{TBC2table} as the starting point for new simple but approximate alignments,
which lead to phenomenologically viable possibilities which differ from either TBC or TBC2 mixing which were used
to discover the new alignments, but which can subsequently be discarded. Thus the alignment 
in Eq.\ref{Phi4} which reproduces TBC mixing 
is discarded in favour of the approximate but simpler alignment in Eq.\ref{Phi3} which gives the different but phenomenologically viable mixing in Eq.\ref{values1}.

As a second example from Table~\ref{TBCtable} we find the following phenomenological possibility,
corresponding to approximate TBC mixing with $\delta =\pi/2$ and $\beta =3\pi/8$,
where this case corresponds to $\eta = \eta_2=\eta_3=\pi/3$, 
$|z_1|=1$, $|z_2|=1$, $|z_3|=3$. 
This suggests the alignments, in our phase convention,
\begin{equation}
\label{Phi5} 
A \equiv  \begin{pmatrix} 0 \\ 1 \\ 1\end{pmatrix},
 \qquad
 B \equiv  \begin{pmatrix}1 \\ 1  \\ 3  \end{pmatrix}e^{i\pi/3}.
\end{equation}
In a realistic model, the alignments may be predicted by some family symmetry for example, but not the value of $\epsilon_{\nu}$ which will depend on right-handed neutrino masses and unknown Yukawa couplings. Therefore we shall regard $\epsilon_{\nu}$ as a free parameter
which may be varied to give different the correct ratio
of physical neutrino masses $m_2/m_3$.
For example, using the above alignment with $\epsilon_{\nu}=0.1$
we find using MPT, $m_2/m_3=0.17$ together with,
\beq
\theta_{12}=34.3^o , \ \ \theta_{13}=8.6^o ,\ \ \theta_{23}=44.3^o,  \ \ \delta = 93^o, \ \ \beta = 72^o,
\label{values2}
\eeq
which is in good agreement with the global fits. 

As a third example from Table~\ref{TBCtable} we find the following phenomenological possibility,
corresponding to approximate TBC mixing with $\delta =\pi/2$ and $\beta =-3\pi/8$,
where this case corresponds to $\eta = \eta_2=\eta_3=-\pi/3$, 
$|z_1|=1$, $|z_2|=3$, $|z_3|=1$. 
This suggests the alignments, in our phase convention,
\begin{equation}
\label{Phi6} 
A \equiv  \begin{pmatrix} 0 \\ 1 \\ 1\end{pmatrix},
 \qquad
 B \equiv  \begin{pmatrix}1 \\ 3  \\ 1  \end{pmatrix}e^{-i\pi/3}.
\end{equation}
For example, using the above alignment with $\epsilon_{\nu}=0.1$
we find using MPT, $m_2/m_3=0.17$ together with,
\beq
\theta_{12}=34.3^o , \ \ \theta_{13}=8.6^o ,\ \ \theta_{23}=45.7^o,  \ \ \delta = 87^o, \ \ \beta = -72^o,
\label{values3}
\eeq
which again is in good agreement with the global fits. 

We emphasise that, with the alignments including the phase $\eta$
fixed, the neutrino mass matrix is completely determined by only two
parameters in Eq.\ref{seesaw2}, namely an overall mass scale $m_a$, which may be taken
to fix the atmospheric neutrino mass $m_3=0.048-0.051$ eV, the ratio 
of input masses $\epsilon_{\nu}$, which may be taken to fix the solar to atmospheric neutrino mass
ratio $m_2/m_3 = 0.17-0.18$. In particular the entire PMNS mixing matrix and all the 
parameters therein are then predicted as a function
of $m_2/m_3$ controlled by the only remaining
parameter $\epsilon_{\nu}$. 
In Table~\ref{predictions142} we show the predictions for the first example
above in Eq.\ref{Phi3} as a function of $\epsilon_{\nu}$ and hence $m_2/m_3$.
In Table~\ref{CSD3predictions113} we show the predictions for the second example
above in Eq.\ref{Phi5} as a function of $\epsilon_{\nu}$ and hence $m_2/m_3$.
In Table~\ref{CSD3predictions} we show the predictions for the third example
above in Eq.\ref{Phi6} as a function of $\epsilon_{\nu}$ and hence $m_2/m_3$.
The results in Table~\ref{CSD3predictions113} and Table~\ref{CSD3predictions}
only differ in the atmospheric angle and the oscillation phase $\delta$,
which are correlated via an atmospheric sum rule discussed later in Eq.\ref{atmsum},
with both sets of results approximated by TBC mixing to within an accuracy of one degree, but with the phases predicted.

We remark that an accuracy of one degree in the angles is all that can be expected due to purely theoretical
corrections in a realistic model due to renormalisation group running \cite{Boudjemaa:2008jf} and
canonical normalisation corrections \cite{Antusch:2007ib}.
In addition, there may be small contributions from a heavy third right-handed neutrino \cite{Antusch:2010tf}
which can affect the results.

\begin{table}
	\centering
		\begin{tabular}{|c||c|c|c|c|c|c|}
			\hline
			 $\epsilon_{\nu}$ & $m_2/m_3$ &  $\theta_{12}$ 
			 & $\theta_{13}$  & $\theta_{23}$  & $\delta$  & $\beta$ \\ \hline \hline
			0.057 &0.166 &34.2$^{\circ}$ & 9.0$^{\circ}$ & 40.8$^{\circ}$ & 107$^{\circ}$ & -84$^{\circ}$\\ \hline   
			0.058 &0.170 &34.2$^{\circ}$ & 9.2$^{\circ}$ & 40.9$^{\circ}$ & 107$^{\circ}$ & -83$^{\circ}$\\ \hline 
			0.059 &0.174 &34.1$^{\circ}$ & 9.4$^{\circ}$ & 41.0$^{\circ}$ & 106$^{\circ}$ & -82$^{\circ}$\\ \hline
			0.060 &0.177 &34.1$^{\circ}$ & 9.6$^{\circ}$ & 41.1$^{\circ}$ & 105$^{\circ}$ & -80$^{\circ}$\\ \hline  
			0.061 &0.181 &34.1$^{\circ}$ & 9.7$^{\circ}$ & 41.3$^{\circ}$ & 104$^{\circ}$ & -79$^{\circ}$\\ \hline     
				\end{tabular} 
			\caption{The predictions for PMNS parameters and $m_2/m_3$
			arising from Eq.\ref{Phi3} as a function of 
			$\epsilon_{\nu}$. Note that these predictions assume $\eta = -2\pi/5$.
			Identical results are obtained for $\eta = 2\pi/5$ with the phases 
			$\beta$ and $\delta$ changed in sign.
			}
		\label{predictions142}
\end{table}

\begin{table}
	\centering
		\begin{tabular}{|c||c|c|c|c|c|c|}
			\hline
			 $\epsilon_{\nu}$ & $m_2/m_3$ &  $\theta_{12}$ 
			 & $\theta_{13}$  & $\theta_{23}$  & $\delta$  & $\beta$ \\ \hline \hline
			0.098 &0.168 &34.4$^{\circ}$ & 8.4$^{\circ}$ & 44.4$^{\circ}$ & 92$^{\circ}$ & -73$^{\circ}$\\ \hline   
			0.100 &0.171 &34.3$^{\circ}$ & 8.6$^{\circ}$ & 44.3$^{\circ}$ & 93$^{\circ}$ & -72$^{\circ}$\\ \hline 
			0.102 &0.173 &34.3$^{\circ}$ & 8.75$^{\circ}$ & 44.1$^{\circ}$ & 94$^{\circ}$ & -71$^{\circ}$\\ \hline
			0.104 &0.177 &34.3$^{\circ}$ & 8.9$^{\circ}$ & 44.0$^{\circ}$ & 94$^{\circ}$ & -70$^{\circ}$\\ \hline  
			0.106 &0.179 &34.2$^{\circ}$ & 9.1$^{\circ}$ & 43.8$^{\circ}$ & 95$^{\circ}$ & -69$^{\circ}$\\ \hline     
				\end{tabular} 
			\caption{The predictions for PMNS parameters and $m_2/m_3$
			arising from CSD3 in Eq.\ref{Phi5} as a function of 
			$\epsilon_{\nu}$. Note that these predictions assume $\eta = \pi/3$.
			Identical results are obtained for $\eta = -\pi/3$ with the phases 
			$\beta$ and $\delta$ changed in sign.
			These angle predictions are approximately equal to the 
			Tri-bimaximal-Cabibbo mixing angle values $\theta_{12}=35.26^{\circ}$, $\theta_{13}=9.15^{\circ}$,
			$\theta_{23}=45^{\circ}$ to within one degree.
			}
		\label{CSD3predictions113}
\end{table}

\begin{table}
	\centering
		\begin{tabular}{|c||c|c|c|c|c|c|}
			\hline
			 $\epsilon_{\nu}$ & $m_2/m_3$ &  $\theta_{12}$ 
			 & $\theta_{13}$  & $\theta_{23}$  & $\delta$  & $\beta$ \\ \hline \hline
			0.098 &0.168 &34.4$^{\circ}$ & 8.4$^{\circ}$ & 45.5$^{\circ}$ & 88$^{\circ}$ & -73$^{\circ}$\\ \hline   
			0.100 &0.171 &34.3$^{\circ}$ & 8.6$^{\circ}$ & 45.7$^{\circ}$ & 87$^{\circ}$ & -72$^{\circ}$\\ \hline 
			0.102 &0.173 &34.3$^{\circ}$ & 8.75$^{\circ}$ & 45.9$^{\circ}$ & 86$^{\circ}$ & -71$^{\circ}$\\ \hline
			0.104 &0.177 &34.3$^{\circ}$ & 8.9$^{\circ}$ & 46.0$^{\circ}$ & 86$^{\circ}$ & -70$^{\circ}$\\ \hline  
			0.106 &0.179 &34.2$^{\circ}$ & 9.1$^{\circ}$ & 46.2$^{\circ}$ & 85$^{\circ}$ & -69$^{\circ}$\\ \hline     
				\end{tabular} 
			\caption{The predictions for PMNS parameters and $m_2/m_3$
			arising from CSD3 in Eq.\ref{Phi6} as a function of 
			$\epsilon_{\nu}$. Note that these predictions assume $\eta = -\pi/3$.
			Identical results are obtained for $\eta = \pi/3$ with the phases 
			$\beta$ and $\delta$ changed in sign.
			These angle predictions are approximately equal to the 
			Tri-bimaximal-Cabibbo mixing angle values $\theta_{12}=35.26^{\circ}$, $\theta_{13}=9.15^{\circ}$,
			$\theta_{23}=45^{\circ}$ to within one degree.
			}
		\label{CSD3predictions}
\end{table}

\section{Leading Order Analytic Results }
\label{LO}
The above results are numerical, based on the exact analytic master formula
in Eq.\ref{master}.
In order to get some feeling for the results it is worth taking a look at the well known but approximate analytic
expressions for the masses and mixing parameters in terms of the Dirac mass matrix elements and 
right-handed neutrino masses, bearing in mind that these results may have large corrections.
The neutrino masses in terms of the parameters in Eq.\ref{2rh1}
are given to leading order in $m_2/m_3$ by \cite{King:2002nf,King:2002qh},
\bea
m_1 & = & 0 \\
m_2 & \approx &  \frac{|a|^2}{Xs_{12}^2}  \label{m2} \\
m_3 & \approx & \frac{|e|^2+|f|^2}{Y} 
\label{m3}
\eea
The neutrino mixing angles are given to leading order in $m_2/m_3$ by
\cite{King:2002nf,King:2002qh},
\bea
\tan \theta_{23} & \approx & \frac{|e|}{|f|} \label{23}\\
\tan \theta_{12} & \approx &
\frac{|a|}
{c_{23}|b|
\cos(\tilde{\phi}_b)-
s_{23}|c|
\cos(\tilde{\phi}_c)} \label{12} \\
\theta_{13} & \approx &
e^{i(\tilde{\phi}+\phi_a-\phi_e)}
\frac{|a|(e^*b+f^*c)}{[|e|^2+|f|^2]^{3/2}}
\frac{Y}{X}
\label{13}
\eea
where we have written some (but not all) complex Yukawa couplings as
$x=|x|e^{i\phi_x}$. The Dirac CP violating oscillation phase $\delta$
is fixed to give a real angle
$\theta_{12}$ by,
\beq
c_{23}|b|
\sin(\tilde{\phi}_b)
\approx
s_{23}|c|
\sin(\tilde{\phi}_c)
\label{chi1}
\eeq
where 
\bea
\tilde{\phi}_b &\equiv & 
\phi_b-\phi_a-\tilde{\phi}+\delta, \nonumber \\ 
\tilde{\phi}_c &\equiv & 
\phi_c-\phi_a+\phi_e-\phi_f-\tilde{\phi}+\delta
\label{bpcp}
\eea
The phase $\tilde{\phi}$
is fixed to give a real angle
$\theta_{13}$ by \cite{King:2002nf},
\beq
\tilde{\phi} \approx  \phi_e-\phi_a -\phi
\label{phi2dsmall}
\eeq
where
\beq
\phi =\arg(e^*b+f^*c). 
\label{lepto0}
\eeq
Eq.\ref{lepto0} may be expressed as
\beq
\tan \phi \approx 
\frac{|b|s_{23}s_2+|c|c_{23}s_3}{|b|s_{23}c_2+|c|c_{23}c_3}.
\label{phi121}
\eeq
Inserting $\tilde{\phi}$ in Eq.\ref{phi2dsmall}
into Eqs.\ref{chi1},\ref{bpcp}, we obtain,
\beq
\tan (\phi+\delta) \approx 
\frac{|b|c_{23}s_2-|c|s_{23}s_3}{-|b|c_{23}c_2+|c|s_{23}c_3}
\label{phi12del}
\eeq
where we have written $s_i=\sin \eta_i, c_i=\cos \eta_i$
where
\beq
\eta_2\equiv \phi_b-\phi_e, \ \ \eta_3\equiv \phi_c-\phi_f
\label{eta}
\eeq
are the phases introduced in Eq.\ref{phibc} which are 
invariant under a charged lepton phase transformation.
The reason that the see-saw parameters only
involve two invariant phases $\eta_2, \eta_3$ rather than the usual six
is due to the two-right handed neutrino assumption, which removes three phases, 
together with the assumption
of a dominant texture zero, which removes another phase.

From the above approximate leading order
results we find the following alternative useful expressions for the mixing angles,
\bea
\tan \theta_{23} & \approx &|z_1| \label{231}\\
\cot \theta_{12} & \approx &
c_{23}|z_2|\cos\left( \eta_2-\frac{\beta}{2} \right)-s_{23}|z_3|\cos \left( \eta_3-\frac{\beta}{2}\right) \label{121}\\
\theta_{13} & \approx & \frac{m_2}{m_3}s_{12}^2c_{23}\left||z_3|  +|z_2|\tan \theta_{23}
e^{i(\eta_2 -\eta_3)}     \right|
\label{131}
\eea
where $\eta_2$ and $\eta_3$ are given in Eq.\ref{phibc} and $|z_i|$ are given in Eq.\ref{ratios2}.
The above approximate leading order results should be compared to the exact results 
in Eqs.\ref{master6}-\ref{master66}. In fact they may be derived by inverting Eqs.\ref{master6}-\ref{master66}.
This is non-trivial to do exactly, but to leading order in $m_2/m_3$ it leads to the results in Eqs.\ref{231}-\ref{131},
using the standard PDG expressions for the mixing elements. We emphasise that these leading order results are very crude, for example we have seen from Fig.\ref{z1} that for close to maximal mixing 
$z_1\approx 1$ has a 50\% variation depending on the phase $\delta$.
Indeed maximal atmospheric mixing prefers $\delta \approx \pi/2$.
This is due to next to leading order effects beyond the simple approximations above.
However the leading order results do provide some understanding of the successful alignments
discussed in the previous subsection as we now discuss.

The alignments discussed in the previous subsection have $\eta = \eta_2=\eta_3$
and lead to approximate maximal mixing with $|z_1|=1$ and $\tan \theta_{23}\approx 1$. 
In this case the approximate formulas above become,
\bea
\cot \theta_{12} & \approx &
 \frac{1}{\sqrt{2}}\cos\left( \eta-\frac{\beta}{2} \right) (|z_2|-|z_3|) \label{122}\\
\theta_{13} & \approx & \frac{s_{12}^2}{\sqrt{2}}\frac{m_2}{m_3}(|z_3|  +|z_2|).
\label{132}
\eea
These approximate results show that the solar angle depends on the difference between $|z_2|$ and $|z_3|$
and the phase difference $\eta-\frac{\beta}{2}$. Phenomenologically we must arrange to obtain $\cot \theta_{12}\approx \sqrt{2}$ leading to the approximate condition for the difference,
\bea
|z_2|-|z_3| & \approx & \frac{2}{\cos(\eta-\frac{\beta}{2} )} \label{123}
\eea
Since $s_{12}^2 \approx 1/3$ the reactor angle is approximately given by,
\bea
\theta_{13} & \approx & \frac{1}{3\sqrt{2}}\frac{m_2}{m_3}(|z_2|  +|z_3|).
\label{133}
\eea
Since phenomenologically ${m_2}/{m_3}\approx \lambda/\sqrt{2}$ and $\theta_{13}\approx \lambda/\sqrt{2}$,
we obtain the approximate condition for the sum,
\bea
|z_2|  +|z_3|\approx 3\sqrt{2}.
\label{134}
\eea

In the examples where $\eta = \eta_2 = \eta_3$, there is only a single see-saw phase $\eta$ and so all
physical phases must be related to it. In particular we find, using the above results
and other approximate results for the phases in \cite{King:2002nf,King:2002qh} that, for $|z_2|>|z_3|$,
$\phi = \eta$ and $\delta \approx -2\eta$
and $\beta \approx 2\eta$.  Thus the Majorana phase $\beta$ is simply related to the oscillation
phase $\delta$ according to $\delta \approx -\beta$.
From Eq.\ref{123} we find in these cases,
\bea
|z_2|-|z_3| & \approx & 2\label{1231}.
\eea
Solving Eqs.\ref{134} and \ref{1231} we find,
\be
|z_2|\approx  \frac{3}{\sqrt{2}} + 1 \approx 3.1, \ \   |z_3|\approx \frac{3}{\sqrt{2}} -1 \approx 1.1.
\label{estimate}
\ee
These results are well respected by the third example in Eq.\ref{Phi6} corresponding to
$|z_1|=1$, $|z_2|=3$, $|z_3|=1$. For the second example analogous results to those in Eq.\ref{estimate}
may be derived but with $|z_2|$ and $|z_3|$ being interchanged, which is in good agreement
with $|z_2|=1$, $|z_3|=3$. The first example in Eq.\ref{Phi3} with 
$|z_1|=1$, $|z_2|=4$, $|z_3|=2$ does not obey the estimate in Eq.\ref{estimate} very well.
Nevertheless it leads to good agreement with experiment, which highlights the limitations of the leading order results,
and the importance of the exact master formula.

\section{Link between Leptogenesis and PMNS phases }
\label{LG}
In the two right-handed neutrino model with dominant texture zero, that we are considering in this paper,
Eq.\ref{phibc2} shows that there are only two see-saw phases that are relevant, namely $\eta_2$ and 
$\eta_3$. The low energy PMNS phases $\delta$ and $\beta$ are therefore (complicated) functions
of these two see-saw phases, and the other see-saw parameters, with the exact relation
obtained via the master formula in Eq.\ref{master}. The phases relevant for leptogenesis
can also only depend on the two see-saw phases $\eta_2$ and $\eta_3$. This implies that the
phases which control leptogenesis 
must be related to the two physical low energy PMNS phases $\delta$ and $\beta$.
This means that by studying CP violation in the laboratory one is in fact also probing
CP violation in the early universe which was relevant for matter-antimatter asymmetry.

The Leptogenesis-PMNS link for the two right-handed neutrino model with dominant texture zero
been discussed some time ago in the literature, first for flavour independent leptogenesis
in \cite{King:2002qh}, then later for flavour dependent leptogenesis in \cite{Antusch:2006cw}.
In the flavour independent leptogenesis case \cite{King:2002qh}
the lepton asymmetry is proportional to $\sin 2\phi$ where 
the relevant leptogenesis phase is $\phi$ was defined 
in Eq.\ref{lepto0}.
In the flavour dependent leptogenesis case \cite{Antusch:2006cw} muon-type lepton asymmetry
is proportional to $\sin (\phi + \eta_2)$,
with tau-type lepton asymmetry being proportional to $\sin (\phi + \eta_3)$
and electron-type lepton asymmetry being identically zero due to the texture zero. 

Estimates of the baryon asymmetry $Y_B$, including the effects of washout, 
for the case where the lightest right-handed neutrino dominates the production of baryon asymmetry
(so-called $N_1$ leptogenesis) for the flavour-dependent case were made in \cite{Antusch:2006cw},
in particular for the above case of the two right-handed neutrino model with a dominant texture zero.
When the lightest right-handed neutrino is identified with $N_{\rm atm}$, one finds \cite{Antusch:2006cw},
assuming $|e|\approx |f|$ which leads to equal washouts $\eta_{\mu}\approx \eta_{\tau}$,
\beq
Y_B\propto |b|\sin (\phi + \eta_2)
+ |c|\sin (\phi + \eta_3).
\label{YB1}
\eeq
When the lightest right-handed neutrino is identified with $N_{\rm sol}$, one finds \cite{Antusch:2006cw},
assuming $|e|\approx |f|$, with unequal washouts in this case,
\beq
Y_B\propto -\eta_{\mu}|b|\sin (\phi + \eta_2)
-\eta_{\tau} |c|\sin (\phi + \eta_3).
\label{YB2}
\eeq
The above estimates assume that the lightest right-handed neutrino dominates the production of baryon asymmetry
($N_1$ leptogenesis). 
In certain situations the heavier right-handed neutrino can dominate
($N_2$ leptogenesis) as was discussed for the two right-handed neutrino model, especially for the case where the lightest right-handed neutrino is identified with $N_{\rm atm}$
\cite{Antusch:2011nz}. In this case the situation is more involved, and it is not possible to give simple
analytic estimates for the baryon asymmetry as above. However we remark that for $N_2$ leptogenesis the 
flavour dependent lepton asymmetries are proportional to the same phase factors as for $N_1$ leptogenesis 
but have opposite sign.

For the examples where $\eta = \eta_2 = \eta_3$ the baryon asymmetry $Y_B$ is clearly proportional to a common 
factor $\sin (\phi + \eta)$. From Eq.\ref{lepto0} we see that in this case $\phi = \eta$ which is an exact result.
Hence the baryon asymmetry is proportional to $\sin (2\eta)$ or $\sin (2\phi)$ which makes sense since
$\eta$ is the only remaining see-saw phase in this case. 
The physical low energy CP violating phases 
$\beta $ and $\delta$ must vanish in the limit that $\eta \rightarrow 0$ since $\eta$ is the only high energy
source of CP violation at the see-saw scale.
The exact relationship between $\beta , \delta$ and $\eta$
provided via the master formula is rather complicated,
however the leading order estimates for the phases derived from \cite{King:2002nf,King:2002qh} are
$\delta \approx -2\eta$
and $\beta \approx 2\eta$, 
which are in fair agreement with the numerical results in Eq.\ref{values3}.
Thus, approximately, we have in this case that the lepton asymmetry is proportional to $\sin \delta$,
which is maximal in this case.
The main message is that, by studying CP violation in oscillation experiments, one is in fact also probing
the source of CP violation in the early universe which was relevant for matter-antimatter asymmetry.

\section{A Minimal Predictive See-Saw Model}
\label{CSD3}

\subsection{CSD3 and vacuum alignment}
In discussing the implications of our results for model building, we have been guided by the idea that the most attractive flavon alignments are the simplest ones, since they may eventually arise from some realistic model.
Although we have tabulated several relatively simple possibilities, we have only explored three 
examples from Table~\ref{TBCtable} in detail. These three examples all satisfy the conditions that 
$\eta \equiv \eta_2=\eta_3$ so that the components of a given flavon all carry the same phase 
and lead to approximate maximal mixing with $|z_1|=1$ which means that the atmospheric 
flavon alignment is particularly simple and retains the CSD form in Eq.\ref{Phi0}.
However the price of the non-zero reactor angle (at least in our minimal approach where we maintain the zero in the atmospheric vacuum alignment, and do not invoke charged lepton corrections) 
is that the solar flavon alignment must differ from the CSD alignment shown in Eq.\ref{Phi0},
and in general will be more complicated. The question then arises as to whether such alignments 
may be achieved theoretically from vacuum alignment methods.

In this subsection we shall discuss two of the successful flavon alignments which we have discovered in our general analysis, and show how they may be derived 
from vacuum alignment methods. The two examples we shall consider 
(which correspond to the second and third examples discussed earlier in this section) 
are summarised below,
\begin{equation}
\label{Phi7} 
\frac{\vev{\phi_{\rm atm}}}{\Lambda}
\propto  \begin{pmatrix}0 \\ 1 \\ 1\end{pmatrix}e^{i\phi_e},
 \qquad
\frac{\vev{\phi_{\rm sol}}}{\Lambda}
\propto  
 \begin{pmatrix}1 \\ 1 \\ 3\end{pmatrix}e^{i\phi_b}
 \ \ 
 {\rm or}
\ \ 
\begin{pmatrix}1 \\ 3 \\ 1\end{pmatrix}e^{i\phi_b},
\end{equation}
These two new alignment possibilities shall both be referred to as CSD3, since they both represent new
possibilities not considered before in the literature.
As in the case of CSD2 \cite{Antusch:2011ic}
the above alignments involve the components of each of the flavon alignments 
$\vev{\phi_{\rm atm}}$ and $\vev{\phi_{\rm sol}}$ being relatively real within each flavon,
although the overall phase of that flavon will in general be non-zero.
However the overall phase difference between the flavon alignments $\eta = \phi_b-\phi_e$ is physically
significant and we have seen that particular values of this phase difference are preferred.
In a particular phase convention, the preferred values are displayed in Eqs.\ref{Phi5} and \ref{Phi6}.

The vacuum alignments associated with TB mixing have been very well studied.
Here we shall focus on the family symmetry $A_4$ as it is the smallest non-Abelian
finite group with an irreducible triplet representation. We apply the $A_4$ basis in
which the triplets are explicitly real as given for example
in~\cite{A4-refs}. We shall work in the basis of \cite{King:2011ab} where, denoting a general $A_4$ triplet as
${\bf{c}}=(c_1,c_2,c_3)^T$ and defining $\omega=e^{2\pi i/3}$, the product
rules can be summarised as 
\be
{\bf{c\otimes c'}}  ~=~ \sum_{r=0}^2 (c^{}_1 c'_1 + \omega^{-r} c^{}_2 c'_2 +
\omega^{r} c^{}_3 c'_3 ) \;+\,
\begin{pmatrix} c^{}_2c'_3 \\ c^{}_3c'_1\\c^{}_1c'_2  
\end{pmatrix}
+
\begin{pmatrix} c^{}_3c'_2 \\ c^{}_1c'_3\\c^{}_2c'_1  
\end{pmatrix} \ , \label{basis3}
\ee
 corresponding to two triplets and the sum of the three one-dimensional
 irreducible representations ${\bf{1_r}}$, with ${\bf{1_0=1}}$ being the trivial
 singlet.

Following the methods of \cite{King:2011ab} it is straightforward to obtain the vacuum 
alignments for charged lepton flavon alignments suitable for a diagonal charged lepton mass matrix.
The charged lepton flavon alignments used to generate a diagonal charged lepton mass
matrix are obtained from the renormalisable superpotential \cite{King:2011ab},
\be
W_{A_4}^{\mathrm{flavon},\ell}~\sim~
A_e \varphi_e \varphi_e + A_\mu \varphi_\mu \varphi_\mu 
+ A_\tau \varphi_\tau \varphi_\tau
+O_{e\mu} \varphi_e  \varphi_\mu+ O_{e\tau} \varphi_e  \varphi_\tau
+ O_{\mu\tau} \varphi_\mu  \varphi_\tau \ .  \label{a4-align-charged}
\ee
The triplet driving fields $A_{e,\mu,\tau}$ give rise to flavon alignments
$\langle \varphi_{e,\mu,\tau} \rangle$
with two zero components, and the singlet driving fields $O_{ij}$ require
orthogonality among the three flavon VEVs so that we arrive at the vacuum
structure \cite{King:2011ab},
\be
\langle \varphi_e \rangle =v_e 
\begin{pmatrix} 1\\0\\0 \end{pmatrix}  \ , \qquad
\langle \varphi_\mu \rangle =v_\mu 
\begin{pmatrix} 0\\1\\0 \end{pmatrix} \ , \qquad
\langle \varphi_\tau \rangle = v_\tau 
\begin{pmatrix} 0\\0\\1 \end{pmatrix} 
 \ .\label{a4-align-char0}
\ee
Of more interest to us in this paper are the new neutrino flavon alignments.
The starting point for the discussion is the usual standard 
TB neutrino flavon alignments proportional to the respective columns of the TB mixing matrix,
\be
\langle \varphi_{\nu_1} \rangle = 
v_{\nu_1} \begin{pmatrix} 
  2\\-1\\1 \end{pmatrix}  , 
\qquad
\langle \varphi_{\nu_2} \rangle = 
v_{\nu_2} \begin{pmatrix} 
  1\\1\\-1 \end{pmatrix}  , 
\qquad
\langle \varphi_{\nu_3} \rangle =  
v_{\nu_3} \begin{pmatrix} 0\\1\\1 \end{pmatrix} .
\label{a4-align-nu}
\ee
In the remainder of this subsection we shall show how to obtain the neutrino flavon alignments including the new
alignment,
\be
\langle \varphi_{\nu_4} \rangle = 
v_{\nu_4} \begin{pmatrix} 
  1\\3\\1 \end{pmatrix} ,
\label{a4-align-nu-4}
\ee
which is an example of a CSD3 solar flavon alignment in Eq.\ref{Phi7}.
In constructing a CSD3 model in the notation of Eq.\ref{Phi7} we shall identify 
$\phi_{\rm atm}\equiv \varphi_{\nu_3}$, and $\phi_{\rm sol}\equiv \varphi_{\nu_4}$.
The renormalisable superpotential involving the driving fields
necessary for aligning the neutrino-type flavons is given as\footnote{The alternative CSD3 alignment 
$\langle \varphi_{\nu_4} \rangle \propto (1,1,3)^T$ may readily be obtained by replacing 
$ \varphi_{\mu} \rightarrow  \varphi_{\tau}$ in Eq.\ref{a4-flavon-nu}.}
\bea
W_{A_4}^{\mathrm{flavon},\nu}&=& 
A_{\nu_2} (g_1 \varphi_{\nu_2}\varphi_{\nu_2}  
+ g_2 \varphi_{\nu_2} \xi_{\nu_2}  )
\label{a4-flavon-nu}
\\[2mm] &&
+ \,O_{e\nu_3} g_3 \varphi_e \varphi_{\nu_3}  +  O_{\nu_2 \nu_3} g_4 \varphi_{\nu_2} \varphi_{\nu_3} 
+  O_{\nu_1 \nu_2} g_5 \varphi_{\nu_1} \varphi_{\nu_2} 
+  O_{\nu_1 \nu_3} g_6 \varphi_{\nu_1} \varphi_{\nu_3} 
\notag \\[2mm] &&
+ \,O_{\mu \nu_5} g_7 \varphi_{\mu} \varphi_{\nu_5} +  O_{\nu_2 \nu_5} g_8 \varphi_{\nu_2} \varphi_{\nu_5}
+ O_{\mu {\nu_6}} g_9 \varphi_{\mu} \varphi_{\nu_6} +  O_{{\nu_5} {\nu_6}} g_{10}\varphi_{\nu_5} \varphi_{\nu_6}
\notag \\[2mm] &&
+\,O_{{\nu_6} \nu_4} g_{11} \varphi_{\nu_6} \varphi_{\nu_4}
+\,O_{\nu_1 \nu_4} g_{12} \varphi_{\nu_1} \varphi_{\nu_4},
\notag
\eea
where $A_{\nu_2}$ is a triplet driving field and $O_{ij}$ are singlet driving fields whose
F-terms lead to orthogonality relations between the accompanying flavon fields.
Here $g_i$ are dimensionless coupling constants.
The first line of Eq.~\eqref{a4-flavon-nu} produces the vacuum alignment 
$\langle \varphi_{\nu_2} \rangle \propto (1,1,-1)^T$ of Eq.~\eqref{a4-align-nu}
as can be seen from the $F$-term conditions\footnote{We remark that the
  general alignment derived from these $F$-term conditions is $\langle
  \varphi_{\nu_2} \rangle \propto (\pm1 ,\pm 1,\pm1)^T$. One can, however,
  show that all of them are equivalent up to phase redefinitions. 
  Note that $(1,1,-1)$ is related to permutations of the minus sign as well as to 
  $(-1,-1,-1)$ by $A_4$ transformations. The other four choices can be obtained from these by simply multiplying an overall phase (which would also change the sign of the $\xi_{\nu_2}$ vev.) } 
\be
2 g_1 \begin{pmatrix}
\langle \varphi_{\nu_2} \rangle_2  \langle \varphi_{\nu_2} \rangle_3 \\
\langle \varphi_{\nu_2} \rangle_3  \langle \varphi_{\nu_2} \rangle_1 \\
\langle \varphi_{\nu_2} \rangle_1  \langle \varphi_{\nu_2} \rangle_2 
\end{pmatrix}
+ g_2 \langle \xi_{\nu_2}  \rangle 
\begin{pmatrix}
\langle \varphi_{\nu_2} \rangle_1\\
\langle \varphi_{\nu_2} \rangle_2\\
\langle \varphi_{\nu_2} \rangle_3
\end{pmatrix} ~=~ 
\begin{pmatrix} 0\\0\\0
\end{pmatrix} . 
\ee

The first two terms in the second line of Eq.~\eqref{a4-flavon-nu} give rise to
orthogonality conditions which uniquely fix the alignment 
$\langle \varphi_{\nu_3} \rangle \propto (0,1,1)^T$ of Eq.~\eqref{a4-align-nu},
\bea
\langle \varphi_e \rangle^T \cdot \langle \varphi_{\nu_3} \rangle \,=\, 
\langle \varphi_{\nu_2} \rangle^T \cdot \langle \varphi_{\nu_3} \rangle \,=\, 0 
\quad 
&\rightarrow&
\quad
\langle \varphi_{\nu_3} \rangle \,\propto\, \begin{pmatrix} 0\\1\\1\end{pmatrix} \ .
\eea

The last two terms in the second line of Eq.~\eqref{a4-flavon-nu} give rise to
orthogonality conditions which uniquely fix the alignment 
$\langle \varphi_{\nu_1} \rangle \propto (2,-1,1)^T$ of Eq.~\eqref{a4-align-nu},
\bea
\langle \varphi_{\nu_1} \rangle^T \cdot \langle \varphi_{\nu_2} \rangle \,=\, 
\langle \varphi_{\nu_1} \rangle^T \cdot \langle \varphi_{\nu_3} \rangle \,=\, 0 
\quad 
&\rightarrow&
\quad
\langle \varphi_{\nu_1} \rangle \,\propto\, \begin{pmatrix} 2\\-1\\1\end{pmatrix} \ .
\eea

The terms in the third line of Eq.~\eqref{a4-flavon-nu} give rise to
orthogonality conditions which uniquely fix the alignments of the auxiliary
flavon fields $\varphi_{\nu_5}$ and $\varphi_{\nu_6}$,
\bea
\langle \varphi_{\mu} \rangle^T \cdot \langle \varphi_{\nu_5} \rangle \,=\, 
\langle \varphi_{\nu_2} \rangle^T \cdot \langle \varphi_{\nu_5} \rangle \,=\, 0 
\quad 
&\rightarrow&
\quad
\langle \varphi_{\nu_5} \rangle \,\propto\, \begin{pmatrix} 1\\0\\1\end{pmatrix} \ , \\
\langle \varphi_{\mu} \rangle^T \cdot \langle \varphi_{\nu_6} \rangle \,=\, 
\langle \varphi_{\nu_5} \rangle^T \cdot \langle \varphi_{\nu_6} \rangle \,=\, 0 
\quad 
&\rightarrow&
\quad
\langle \varphi_{\nu_6} \rangle \,\propto\, \begin{pmatrix} 1\\0\\-1\end{pmatrix} \ .
\eea

The neutrino-type flavon of interest labelled as $\varphi_{\nu_4}$ gets aligned
by the remaining terms in the fourth line of Eq.~\eqref{a4-flavon-nu}.
A vanishing $F$-term of
the driving field $O_{{\nu_6} \nu_4}$ requires
\be
\langle \varphi_{\nu_4} \rangle = \begin{pmatrix} n_1 \\ n_2 \\ n_1 \end{pmatrix} ,
\ee
where $n_1$ and $n_2$ are independent parameters. 
Finally a vanishing $F$-term of
the driving field $O_{\nu_1 \nu_4}$ leads to the desired alignment in Eq.\ref{a4-align-nu-4},
\bea
\langle \varphi_{\nu_1} \rangle^T \cdot \langle \varphi_{\nu_4} \rangle \,=\, 0 
\quad 
&\rightarrow&
\quad
\langle \varphi_{\nu_4} \rangle \,\propto\, \begin{pmatrix} 1\\3\\1\end{pmatrix} \ .
\eea

So far we have only shown how to align the flavon vevs and have not enforced them to be non-zero.
In order to do this we shall introduce the 
additional non-renormalisable superpotential terms which include,
\bea
\Delta W_{A_4}^{\mathrm{flavon},\ell}&\sim &  \sum_{l=e,\mu , \tau} \frac{P}{\Lambda} \left( (\varphi_{l} \cdot \varphi_{l} ) \rho_{l} - M^3 \right)
+ \frac{P}{\Lambda} (\rho_{l}^3  - M^3) , \label{Delta-a4-align-charged}
     \\
\Delta W_{A_4}^{\mathrm{flavon},\nu} & \sim &   \sum_{i=1}^6 \frac{P}{\Lambda} \left( (\varphi_{\nu_i}) \cdot \varphi_{\nu_i} ) \rho_{\nu_i} - M^3 \right)  
+ \frac{P}{\Lambda} (\rho_{\nu_i}^3  - M^3) , \label{Delta-a4-flavon-nu}\\
\Delta W_{A_4}^{\mathrm{flavon},\xi_{\nu_2} } & \sim &     \frac{P}{\Lambda} (\xi_{\nu_2} ^3  - M^3) \label{Delta-xi}  \;,
\eea
where, as in \cite{Antusch:2011sx}, the driving singlet fields $P$ denote linear combinations 
of identical singlets and we have introduced explicit masses $M$
to drive the non-zero vevs, as well as the messenger scales denoted as $\Lambda$.
For example the $A_4$ singlet $\xi_{\nu_2} $ has a vev of order $M$ driven by the F-term of the singlet $P$
in Eq.\ref{Delta-xi}.
We have also introduced $A_4$ singlets $\rho_{l}$ and $\rho_{\nu_i}$ whose vevs are driven by the 
F-terms of the singlets $P$ in the second terms in Eqs.\ref{Delta-a4-align-charged}
and \ref{Delta-a4-flavon-nu}. These singlet vevs enter 
the first terms in Eqs.\ref{Delta-a4-align-charged}
and \ref{Delta-a4-flavon-nu} which drive the vevs of the triplet flavons.

The flavons and driving fields introduced in this subsection 
transform under $Z_3^{l}\times Z_3^{\nu_i}$ symmetries
whose purpose is to allow only the terms
in Eqs.\ref{a4-align-charged}, \ref{a4-flavon-nu}
and \ref{Delta-a4-align-charged}-\ref{Delta-xi} and forbid all other terms.
Any superfield with a single subscript $\nu_i$ transforms 
under $Z_3^{\nu_i}$ as $\omega^2$ and is a singlet under all other discrete symmetries. Any superfield with a single subscript $l$ transforms under $Z_3^{l}$ as $\omega$ and is a singlet under all other discrete symmetries.
The orthogonality driving superfields $O_{ij}$ with two subscripts transform under $Z_3^{l}\times Z_3^{\nu_i}$
in such a way as to allow the terms in Eqs.\ref{a4-align-charged}, \ref{a4-flavon-nu}.
For example the  $O_{l {\nu_i}}$ driving fields transform under $Z_3^{l}\times Z_3^{\nu_i}$
as $(\omega^2,\omega)$. In addition driving superfields are assigned a charge of two 
while flavon superfields have zero charge under a $U(1)_R$ symmetry.

\subsection{A predictive $A_4\times Z_3^{10}$ model of leptons }
In this subsection we outline a supersymmetric (SUSY) $A_4$ model of leptons with CSD3
along the lines of the $A_4$ models of leptons discussed in 
\cite{Antusch:2011ic,King:2011ab}. The basic idea is that the three families of lepton doublets
$L$ form a triplet of $A_4$ while the right-handed charged leptons
$e^c, \mu^c, \tau^c$, right-handed neutrinos $N_{\rm atm},N_{\rm sol}$ 
and the two Higgs doublets $H_1,H_2$ required by SUSY are all singlets of $A_4$.
In addition the model employs an additional discrete symmetry 
$Z_3^{l}\times Z_3^{\nu_i}$ introduced in the previous subsection together with a $Z_3^{\theta}$ family symmetry
in order to account for the charged lepton mass hierarchy.
The full discrete group of the model is $A_4\times Z_3^{10}$
which has order 42. 

In Table~\ref{tab-A4} we have displayed the symmetries and superfields relevant for the Yukawa sector and have written $Z_3^{\rm atm}\equiv Z_3^{\nu_3}$
and $Z_3^{\rm sol}\equiv Z_3^{\nu_4}$. The transformation properties 
of the remaining superfields under $Z_3^{l}\times Z_3^{\nu_i}$ was discussed at the end
of the previous subsection and are consistent with the charges shown in Table~\ref{tab-A4}.


The charged lepton sector of the model employs the $A_4$ triplet flavons $\varphi_e,  \varphi_{\mu}, \varphi_{\tau}$
whose alignment was discussed in the previous subsection.
With the lepton symmetries in the upper left of Table~\ref{tab-A4} 
we may enforce the following charged lepton Yukawa superpotential at leading order
\begin{equation}
\mathcal{W}^e_{\text{Yuk}} \sim \frac{1}{\Lambda}H_1 (\varphi_{\tau} \cdot L )\tau^c 
+   \frac{1}{\Lambda^2}  \theta H_1 (\varphi_{\mu}\cdot L) \mu^c +
\frac{1}{\Lambda^3}  \theta^2 H_1( \varphi_e \cdot L) e^c,
\end{equation}
which give the charged lepton Yukawa couplings after the flavons develop their
vevs. $\Lambda$ is a generic messenger mass scale, but in a renormalisable model
the messengers scales may differ. The charged lepton symmetries include 
three lepton flavour symmetries $Z_3^{e, \mu , \tau}$ under which 
$\varphi_e,  \varphi_{\mu}, \varphi_{\tau}$ and $e^c, \mu^c , \tau^c$ 
transform respectively 
as $\omega$ and $\omega^2$, together with a lepton family symmetry $Z_3^{\theta}$ under which 
$e^c, \mu^c , \tau^c$ transform as $\omega , \omega^2 , 1$ respectively
(where $\omega = e^{i2\pi /3}$ ) with the 
family symmetry breaking flavon $\theta$ transforming as $\omega$ and otherwise being a singlet
under all other symmetries. $H_1$ and $L$ and all other fields 
are singlets under $Z_3^{e, \mu , \tau}$ and $Z_3^{\theta}$.
With these charge assignments the higher order corrections are very suppressed.

The charged lepton Yukawa
matrix is diagonal at leading order due to the alignment of the charged lepton-type flavons in Eq.\ref{a4-align-char0}
(where the driving fields responsible for the alignment in Eq.\ref{a4-align-charged} absorb
the charges under the newly introduced symmetries $Z_3^{e, \mu , \tau}$ and $Z_3^{\theta}$)
and has the form, 
\begin{equation}
\label{Ye}
Y^e = \text{diag}(y_e, y_\mu, y_\tau) \sim \text{diag}(\epsilon^2 , \epsilon , 1)y_\tau
\end{equation}
where we choose $\epsilon \sim \langle \theta \rangle /\Lambda \sim \lambda^2$ in order to generate
the correct order of magnitude charged lepton mass hierarchy, with precise
charged lepton masses also dependent on order one coefficients which we have suppressed here.

\begin{table}
	\centering
$$
\begin{array}{||c||cccccccc||c||ccccccc||}
\hline \hline
&\theta &e^c&\mu^c & \tau^c
&\varphi_e&\varphi_\mu&\varphi_\tau
&H_{1}&L&H_{2}&\phi_{\rm atm}&\phi_{\rm sol}
&N_{\rm atm}&N_{\rm sol}
&\xi_{\rm atm}&\xi_{\rm sol}\\  \hline
\hline
Z_3^{\theta} & \omega &  \omega & \omega^2 &1 & 1  &1
&1&1
&1&1 &1
&1& 1  &1&1&1\\[2mm] 
Z_3^{e} & 1 &  \omega^2   & 1  & 1 &  \omega & 1& 1 & 1 & 1 & 1 & 1 
& 1 & 1 & 1  &1 & 1 \\[2mm] 
Z_3^{\mu} & 1 &  1   &   \omega^2  & 1 & 1 &  \omega & 1 & 1
& 1 & 1  & 1  &  1 & 1  & 1 & 1 & 1 \\[2mm] 
Z_3^{\tau} & 1  & 1  & 1 &   \omega^2 & 1 & 1
&  \omega & 1 & 1&  1 & 1 
& 1 & 1  &  1 & 1& 1\\[2mm] \hline \hline
A_4 & {\bf 1} & {\bf 1} & {\bf 1} & {\bf 1} & {\bf 3} & {\bf 3} 
& {\bf 3}  & {\bf 1}
&{\bf 3} & {\bf 1} & {\bf 3} 
& {\bf 3} & {\bf 1}& {\bf 1}& {\bf 1}& {\bf 1} 
\\[2mm] \hline  \hline
Z_3^{\rm atm} & 1 & 1  & 1 & 1&1&1
&1&1
&1& 1 &\omega^2
&1& \omega  &1 &\omega &1\\[2mm] 
Z_3^{\rm sol} & 1 & 1  & 1 & 1&1&1
&1&1
&1& 1 &1
&\omega^2&1 & \omega  &1 &\omega \\[2mm] \hline \hline
%
%
\end{array}
$$
\caption{\label{tab-A4}Lepton, Higgs and flavon superfields 
and how they transform under the symmetries relevant for the Yukawa sector of the model.
The only non-trivial charged lepton charges are in the upper left of the Table
and the only non-trivial neutrino charges in the lower right of the Table.
Note that the only
the lepton doublets $L$ and $A_4$ symmetry, are common to both 
charged lepton and neutrino sectors and are given near the central column and row.
The Standard Model gauge symmetries and $U(1)_R$ symmetry,
under which all the leptons have a charge of unity while the Higgs and flavons have zero charge,
are not shown in the Table. }
\end{table}

The neutrino sector of the model exploits 
the $A_4$ triplet flavons $\phi_{\rm atm}\equiv \varphi_{\nu_3}$, and $\phi_{\rm sol}\equiv \varphi_{\nu_4}$
whose alignment was discussed in the previous subsection.
We shall consider one of the types of CSD3 corresponding to,
\begin{equation}
\label{Phi8} 
\frac{\vev{\phi_{\rm atm}}}{\Lambda}
\propto  \begin{pmatrix}0 \\ 1 \\ 1\end{pmatrix}\equiv A,
 \qquad
\frac{\vev{\phi_{\rm sol}}}{\Lambda}
\propto  
\begin{pmatrix}1 \\ 3 \\ 1\end{pmatrix}\equiv B.
\end{equation}
With the neutrino symmetries in the lower right part of Table~\ref{tab-A4} 
we may enforce the following leading order neutrino Yukawa superpotential 
\begin{equation}
\mathcal{W}^{\nu}_{\text{Yuk}} \sim \frac{1}{\Lambda}H_2 (\phi_{\rm atm} \cdot L) N_{\rm atm}  
+ \frac{1}{\Lambda}H_2 (\phi_{\rm sol} \cdot L) N_{\rm sol}  .
\end{equation}
Again the higher order corrections are completely negligible.
Inserting the vacuum alignments in Eq.\ref{Phi8} the neutrino Yukawa matrix is of the form
\begin{equation} \label{eq:NeutrinoYukawas}
Y^\nu = \begin{pmatrix} 0 & b \\ e & 3 b \\ e & b \end{pmatrix}.
\end{equation}
The parameters $e$ and~$b$ can be determined from the parameters
in the superpotential.

As is typical in models of this kind \cite{Antusch:2011ic,King:2011ab}, the RH neutrinos have no mass terms at the
renormalisable level, but they become massive after 
some $A_4$ singlet flavons $\xi_{\rm atm}$ and 
$\xi_{\rm sol}$ develop their
vevs due to the renormalisable superpotential, 
\begin{equation}
 \mathcal{W}_R \sim \xi_{\rm atm} N_{\rm atm}^2 
 + \xi_{\rm sol} N_{\rm sol}^2  \;.
\end{equation} 
When the 
right-handed neutrino flavons develop their vevs 
$\langle \xi_{\rm atm} \rangle \sim M_A$
together with $\langle \xi_{\rm sol} \rangle \sim M_B$, then the RH neutrino
mass matrix is diagonal as required,
\begin{equation} \label{eq:MR}
 M_R  =  \begin{pmatrix} M_{A} & 0 \\ 0 & M_{B} \end{pmatrix} \;.
\end{equation}
To ensure that the mixed terms are absent at renormalisable order 
we have imposed a right-handed neutrino flavour symmetry
$Z_3^{\rm atm}$ under which $N_{\rm atm}$ and $\xi_{\rm atm}$ both transform as $\omega$
(where $\omega = e^{i2\pi /3}$) while $\phi_{\rm atm}$ transforms as  $\omega^2$,
with all other fields being singlets. We have also imposed a similar symmetry $Z_3^{\rm sol}$
under which the ``solar'' fields transform in an analogous way. 
We remark that these charge assignments are 
consistent with the flavon superpotential in Eq.\ref{a4-flavon-nu}, where we identify 
$\phi_{\rm atm}\equiv \varphi_{\nu_3}$, and $\phi_{\rm sol}\equiv \varphi_{\nu_4}$,
with suitable charges assigned to the driving fields.

The above charge assignments allow higher order non-renormalisable mixed terms 
such as 
\begin{equation}
 \Delta \mathcal{W}_R \sim \frac{1}{\Lambda} (\phi_{\rm atm} . \phi_{\rm sol}) N_{\rm atm} N_{\rm sol}  \;,
\end{equation} 
which contribute off-diagonal terms to the right-handed neutrino mass matrix
of a magnitude which depends on the absolute scale of the 
flavon vevs $\vev{\phi_{\rm atm}}$ and $\vev{\phi_{\rm sol}}$ compared to 
$\langle \xi_{\rm atm} \rangle $ and $\langle \xi_{\rm sol} \rangle $.
If all flavon vevs and messenger scales are set equal then these terms are suppressed by $\epsilon \sim \lambda^2$
according to the estimate below Eq.\ref{Ye}, however they may be even more suppressed.
We shall ignore the contribution of such off-diagonal mass terms in the following.

The effective neutrino mass matrix after implementing the see-saw mechanism can be written 
from Eqs.\ref{seesaw2},\ref{Phi8},\ref{eq:MR},
\beq
m^{\nu}=m_aA^TA+m_ae^{2i\eta}\epsilon_{\nu}B^TB 
= m_a \begin{pmatrix} 0 & 0 & 0 \\ 0 & 1 & 1 \\ 0 & 1 & 1 \end{pmatrix} 
+ m_a e^{2i\eta}\epsilon_{\nu}\begin{pmatrix} 1 & 3 & 1 \\ 3 & 9 & 3 \\ 1 & 3 & 1  \end{pmatrix}
\label{seesaw3}
\eeq
where $m_a$ and $\epsilon_{\nu}$ are real mass parameters which determine the physical neutrino masses 
$m_3$ and $m_2$ and the phase difference $\eta$ was defined in Eq.\ref{eta0}.
Thus the neutrino mass matrix is determined by three input parameters: $m_a$, $\epsilon_{\nu}$ and  $\eta$.
These three input parameters determine nine physical observables consisting of the PMNS mixing matrix plus the three neutrino masses, making this a highly predictive scheme. 
We saw earlier that a relative phase difference $\eta  = -\pi /3$, 
which translates into a Dirac CP phase $\delta \approx \pi /2$ gives a successful set of mixing parameters. 

Such a phase difference $\eta = -\pi /3$ 
could be achieved in the context of spontaneous CP violation from discrete symmetries as discussed in 
\cite{Antusch:2011sx}, and we shall follow the strategy outlined there.
The basic idea is to impose CP conservation on the theory so that all couplings and masses are real.
Note that the $A_4$ assignments in Table~\ref{tab-A4} do not involve the complex
singlets $1',1''$ or any complex Clebsch-Gordan coefficients so that the definition of CP is 
straightforward in this model and hence CP may be defined in different ways which are equivalent for our purposes 
(see \cite{Antusch:2011sx} for a discussion of this point).
The CP symmetry is broken in a discrete way by the form of the superpotential terms.
We shall follow \cite{Antusch:2011sx} and suppose that the flavon vevs  
$\vev{\phi_{\rm atm}}$ and $\vev{\phi_{\rm sol}}$ to be real with the phase $\eta$ in Eq.\ref{seesaw3}
originating from the solar right-handed neutrino mass due to the flavon vev 
$\langle \xi_{\rm sol} \rangle \sim M_Be^{2i\pi/3}$
having a complex phase of $2\pi /3$, while the flavon vev $\langle \xi_{\rm atm} \rangle \sim M_A$ is real and positive. This can be arranged if the right-handed neutrino flavon vevs arise from the superpotential,
\be
W_{A_4}^{\mathrm{flavon},R} = g P\left(\frac{\xi_{\rm atm}^3}{\Lambda}  -M^2\right) + 
g'P' \left(\frac{\xi_{\rm sol}^3}{\Lambda'}  - M'^2\right) ,
\label{Rflavon}
\ee
where, as in \cite{Antusch:2011sx}, the driving singlet fields $P,P'$ denote linear combinations 
of identical singlets and all couplings and masses are real due to CP conservation.
The F-term conditions from Eq.\ref{Rflavon} are,
\begin{equation}
 \left| \frac{\langle \xi_{\rm atm} \rangle^3}{\Lambda} - M^2\right|^2 
 = \left| \frac{\langle \xi_{\rm sol} \rangle^3}{\Lambda'} - M'^2\right|^2 = 0 .
\end{equation} 
These are satisfied by $\langle \xi_{\rm atm} \rangle = |(\Lambda M^2)^{1/3}|$ and 
$\langle \xi_{\rm sol} \rangle =  |(\Lambda'M'^2)^{1/3}|e^{2i\pi/3}$
where we arbitrarily select the phases to be 
zero and $2\pi /3$ from amongst a discrete set of possible choices
in each case. More generally we require a phase difference of $2\pi /3$ since the overall phase is not physically
relevant, which would happen one in three times by chance.
In the basis where the right-handed neutrino masses are real and positive
this is equivalent to $\eta  = -\pi /3$ in Eq.\ref{seesaw3} due to the see-saw mechanism.

Similarly the flavons appearing in Eqs.\ref{Delta-a4-align-charged}-\ref{Delta-xi} each have
a discrete choice of phases $(0,2\pi /3, 4\pi /3)$. The charged lepton flavons $\varphi_l$ may take any phases
since such phases are unphysical. In fact the only physically significant flavon phases from the previous subsection
are those of 
$\phi_{\rm atm}\equiv \varphi_{\nu_3}$, and $\phi_{\rm sol}\equiv \varphi_{\nu_4}$
whose phases are selected to be equal to maintain $\eta  = -\pi /3$ in Eq.\ref{seesaw3}.
As before, this would occur one in three times by chance.

In Table~\ref{CSD3predictions} we show the predictions for this model
corresponding to Eq.\ref{Phi6} fixing $\eta =-\pi /3$
as a function of $\epsilon_{\nu}$ and hence $m_2/m_3$.
With $\eta =- \pi /3$, the PMNS parameters are all accurately predicted
and only depend on the physical neutrino mass ratio $m_2/m_3$ with
$m_1=0$. 
This level of predictiveness for CSD3 is equal to that for CSD2. However in the case of CSD2 the reactor angle
was wrongly predicted, whereas CSD3 reproduces the observed value very well.

As in the case of CSD2, the neutrino mass matrix implies the TM$_1$ 
mixing form \cite{Lam:2006wm} where the first column of the PMNS matrix is proportional to
$(2,-1,1)^T$. The reason is simply that $\langle \varphi_{\nu_1} \rangle \propto (2,-1,1)^T$
is an eigenvector of $m^{\nu}$ in Eq.\ref{seesaw3} with a zero
eigenvalue corresponding to the first neutrino mass $m_1$ being zero. The reason for
this is that $m^{\nu}$ in Eq.\ref{seesaw3} is a sum of two terms, the
first being proportional to $AA^T\propto \langle \varphi_{\nu_3} \rangle \langle \varphi_{\nu_3} \rangle^T$  and
the second being proportional to $BB^T\propto \langle \varphi_{\nu_4} \rangle \langle \varphi_{\nu_4} \rangle^T$.  
Since $\langle \varphi_{\nu_1} \rangle \propto (2,-1,1)^T$ is orthogonal to both
$\langle \varphi_{\nu_3} \rangle$ and $\langle \varphi_{\nu_4} \rangle$ it is then clearly annihilated by
the neutrino mass matrix, i.e. it is an eigenvector with zero
eigenvalue. Therefore we immediately expect $m^{\nu}$ in Eq.\ref{seesaw3}
to be diagonalised by the TM$_1$ mixing matrix \cite{Lam:2006wm} where
the first column is proportional to $\langle \varphi_{\nu_1} \rangle \propto (2,-1,1)^T$. 
Therefore we already know that CSD3
must lead to TM$_1$ mixing exactly to all orders according to this general argument.

Exact  TM$_1$ mixing angle and phase relations are obtained by
equating moduli of PMNS elements to those of the first column of the TB mixing matrix:
\bea
c_{12}c_{13}=\sqrt{\frac{{2}}{{3}}} ,\label{a1} \\
|c_{23}s_{12}+s_{13}s_{23}c_{12}e^{i\delta}|=\frac{1}{\sqrt{6}} , \label{a2}\\
|s_{23}s_{12}-s_{13}c_{23}c_{12}e^{i\delta}|=\frac{1}{\sqrt{6}} . \label{a3}
\eea
From Eq.\ref{a1} we see that TM$_1$ mixing approximately preserves the successful TB mixing for the solar
mixing angle $\theta_{12}\approx 35^\circ$ as the correction due to a non-zero
but relatively small reactor angle is of second order. 
While general TM$_1$ mixing involves an undetermined reactor
angle $\theta_{13}$, we emphasise that CSD3 fixes this reactor angle.
The approximate leading order result from Eq.\ref{133} is
\be
\theta_{13}  \approx  \frac{4}{3\sqrt{2}}\frac{m_2}{m_3}
\label{1333}
\ee
which explains why the bound $\theta_{13}\simlt m_2/m_3$ is approximately saturated.
However the leading order results are not highly accurate and numerically the prediction for the reactor angle
depends on the phase $\eta$. For $\eta = \pi/3$ the reactor angle is in the correct range as shown
in Table~\ref{CSD3predictions}.

In an approximate linear form, the relations in Eq.\ref{a1}-\ref{a3} imply the atmospheric sum rule relation,
\beq
\theta_{23}\approx 45^\circ +\sqrt{2}\theta_{13}\cos \delta .
\label{atmsum}
\eeq
Note that the atmospheric sum rule with 
accurately maximal atmospheric mixing implies maximal CP violation with $\delta \approx  \pi/2$ as we observed
numerically. On the other hand, non-maximal atmospheric mixing is linked to non-maximal CP violation.
For $\eta = \pi/3$ the predictions shown in Table~\ref{CSD3predictions} for the small deviations
of the atmospheric angle from maximality are well described by the sum rule in Eq.\ref{atmsum}.
Clearly the atmospheric
sum rule can be tested in the proposed future high precision neutrino experiments.

Using the deviation parameters in Eq.\ref{rsadef}, the above results may be expressed as,
\beq
s \approx 0, \ \ \ \ r\approx \frac{4}{3}\frac{m_2}{m_3} \approx \lambda,
\ \ \ \ \delta \approx \pm \pi/2,  \ \ \ \ a \approx r\cos \delta ,
\label{TM1r}
\eeq
corresponding to approximate TBC mixing.
Note the novel connection between the Wolfenstein parameter and the ratio of neutrino masses,
which, if postulated to be an exact relation,
\beq
\lambda = \frac{4}{3}\frac{m_2}{m_3},
\eeq
would predict $m_2/m_3 \approx 0.17$.

The TBC mixing matrix in Eq.\ref{TBC}
with $\delta \approx  \pm \pi/2$ becomes,
\beq
V_{\mathrm{TBC}} \approx
\left( \begin{array}{ccc}
\sqrt{\frac{2}{3}}(1-\frac{1}{4}\lambda^2)  & \frac{1}{\sqrt{3}}(1-\frac{1}{4}\lambda^2) 
& \mp \frac{i}{\sqrt{2}}\lambda  \\
-\frac{1}{\sqrt{6}}(1\pm i\lambda )  & \frac{1}{\sqrt{3}}(1\mp \frac{i}{2}\lambda )
& \frac{1}{\sqrt{2}}(1-\frac{1}{4}\lambda^2) \\
\frac{1}{\sqrt{6}}(1\mp i\lambda )  & -\frac{1}{\sqrt{3}}(1\pm \frac{i}{2}\lambda )
 & \frac{1}{\sqrt{2}}(1-\frac{1}{4}\lambda^2)
\end{array}
\right)+ \mathcal{O}(\lambda^3),
\label{TBC3}
\eeq
which is then completely determined in terms of $\lambda$ with no free parameters. 

\section{Conclusions}
\label{conclusion}

The type I see-saw mechanism 
provides a beautiful understanding of the smallness of neutrino masses
as being due to the heavy right-handed Majorana neutrino masses.
However, despite its attractive features, the see-saw mechanism provides 
no understanding of the observed approximate mixing, 
characterised here as TBC or TBC2 mixing.
It also provides no insight into 
either the ordering (i.e. normal or inverted) or the mass scale of the neutrinos (i.e. the mass of the lightest neutrino).
Moreover the see-saw mechanism is difficult to test experimentally, 
and typically contains more parameters than physical observables.
Finally, apparently ``unnatural'' cancellations can take place when constructing the effective neutrino mass matrix from the see-saw parameters.

One attractive idea which avoids the last problem of ``unnatural'' cancellations in the see-saw mechanism
is that the right-handed neutrinos contribute sequentially to the light effective neutrino mass matrix
with hierarchical strength, leading to the prediction
of a normal mass hierarchy of physical neutrino masses $m_3\gg m_2\gg m_1$.
This is consistent with recent {\it Planck} results which provide no evidence of quasi-degeneracy. 
The two right-handed neutrino model then emerges as the limiting case of 
a three right-handed neutrino model with sequential dominance in the limit that $m_1\rightarrow 0$. 

We have considered the minimal type I see-saw model with normal neutrino mass hierarchy
consisting of two right-handed neutrinos with
a zero Yukawa coupling of the ``dominant'' right-handed neutrino to the electron neutrino.
The number of observables is precisely equal to the number of see-saw parameters in this case.
It then becomes possible to derive a new and exact ``master formula'' for such a minimal see-saw model which relates see-saw parameters to physical neutrino observables. We hence 
showed that sequential dominance  
follows automatically and 
a second texture zero is excluded. Using the ``master formula'' we then explored simple ratios of Yukawa couplings in both the dominant and subdominant columns which could explain why the bound on the reactor angle
$\theta_{13}\simlt m_2/m_3$ is saturated by the measurement of $\theta_{13}\sim 0.15$
by Daya Bay and RENO.

The results in Fig.\ref{z1} show that the ratio of Yukawa couplings $|z_1|$ in the dominant column
need not be precisely equal to unity, but may vary from $|z_1|=0.5-1.5$, depending particularly on the CP violating
phase $\delta$. The simple ratio of unity seems to be associated with 
maximal oscillation phase $\delta \approx \pi/2$ for TBC which is an interesting result of the analysis. 
Fixing $\delta= \pi/2$ and $|z_1|=1$, the results in Fig.\ref{TBCdeltapiover2} for TBC mixing then show that 
simple ratios of subdominant Yukawa couplings $|z_2|$ and $|z_3|$ and invariant see-saw phases
$\eta_2$ and $\eta_3$ are also possible for particular values of the Majorana phase $\beta$, with 
some examples shown in Table~\ref{TBCtable}. Analogous results for TBC2 mixing allow some new possibilities
with $\delta= 0$ and $|z_1|=0.5$ which are displayed in Fig.\ref{TBC2deltazero} and Table~\ref{TBC2table}.

From the point of view of model building we have focussed on the simplest possible examples 
of ratios of Yukawa couplings and see-saw phases, later explaining their origin in terms of 
an underlying $A_4$ family symmetry broken by flavons with simple
vacuum alignments. We have been particularly interested in examples where the two
see-saw phases are equal, namely $\eta = \eta_2=\eta_3$ since it is simpler to justify 
vacuum alignments when the components of a particular flavon have equal phases.
In this case then there is only a single input see-saw phase $\eta$,
which controls both leptogenesis and the two PMNS phases $\delta$ and $\beta$,
providing a direct link between cosmological CP violation responsible for matter-antimatter asymmetry,
and the CP violation measured in neutrino oscillation experiments. This appears to be a new 
possibility not discussed before, since people hitherto have focussed on the case
of this link being provided by having two texture zeroes with two right-handed neutrinos.
It would also be worth studying flavour dependent $N_1$ and $N_2$
leptogenesis in detail for such examples along the lines of \cite{Antusch:2011nz}.

We have discovered two particularly simple possibilities based on the atmospheric
flavon alignment $(0,1,1)$ and a solar flavon alignment $(1,1,3)$ or $(1,3,1)$,
involving a relative phase $\eta = \mp \pi/3$, which we call CSD3.
We emphasise the high degree of predictivity
of CSD3 which involves only two input mass parameters to fix 
the three neutrino masses, three mixing angles, one Dirac CP phase
and two Majorana phases (nine physical observables).
We have also seen that the baryon asymmetry is proportional to $\sin (2\eta)$,
which is the only see-saw phase for CSD3, providing a link
between matter-antimatter asymmetry in leptogenesis and low energy CP violation
in neutrino oscillation experiments.
We have checked numerically that CSD3 provides a good description
of PMNS mixing parameters. 
Fixing the relative phase $\eta = \mp \pi/3$ 
by spontaneous CP violation, 
the parameters of PMNS matrix then only depend on the neutrino mass
ratio $m_2/m_3$ as shown in Table~\ref{CSD3predictions} for 
the solar flavon alignment $(1,3,1)$.  
We find it remarkable that such a simple scheme can lead to such a successful
description of all current neutrino data, 
corresponding to approximate TBC mixing to within 
an accuracy of one degree, together with predictions for the phases
$\delta \approx \pm \pi /2$ and $\beta \approx \mp 70^\circ$. 

With CSD3 defined as in Eq.\ref{Phi6} we have an example of a minimal predictive
see-saw model with normal neutrino mass hierarchy.
There are many ways to test the model by the forthcoming neutrino
experiments and it is worth going through them. One way to exclude the model is to 
measure the mixing angles to high precision since according to Table~\ref{CSD3predictions}
these lie in very restricted ranges close (but not quite equal) to the TBC ansatz.
Note the dependence of the reactor angle on the neutrino mass ratio $m_2/m_3$,
which provides another high precision test of the model.
Since the model predicts TM$_1$ mixing, the deviations are in fact predicted analytically.
The solar angle is related to the reactor angle
by the TM$_1$  relation in Eq.\ref{a1}, leading to a value of $\theta_{12}\approx 34^\circ$.
The TM$_1$ atmospheric sum rule in Eq.\ref{atmsum} also leads to an analytic
understanding of the smallness of the deviations of the atmospheric angle from maximality
due to the prediction of the approximately maximal oscillation phase $\delta \approx \pm \pi/2$.
Thus, if the atmospheric angle is determined to be non-maximal, then the model
will be excluded. It is also worth recalling that this is a
two right-handed neutrino model with a normal neutrino mass hierarchy, and so the lightest 
neutrino mass is predicted to be zero $m_1= 0$.
This implies that one of the Majorana phases 
is irrelevant and may be taken to be zero, while the remaining Majorana phase is predicted to be 
$\beta \approx 70^\circ$, although this is hard to measure since the neutrinoless double
beta becay observable is predicted to be very small, 
$m_{\beta \beta}\approx |2.6+1.15e^{-i(\beta +2\delta)}|.10^{-3}\ {\rm eV} \approx 2.5.10^{-3}\ {\rm eV}$.
Nevertheless this is significant since the observation of neutrinoless double beta decay in the near
future would exclude this model, as would any cosmological indication of partially degenerate neutrinos.

We have seen that CSD3 can be realised in an indirect model in which some family symmetry is completely broken
in the neutrino sector, rather than a direct model where some subgroup of the family symmetry 
is preserved in the neutrino sector. We can understand why this is the case by recalling that the neutrino mass matrix 
predicted by CSD3 in Eq.\ref{seesaw3} depends on two mass parameters which determine the 
non-zero neutrino masses and these are two of the input parameters of the model.
Therefore the Klein symmetry of the neutrino mass matrix cannot depend on any of the generators
of the family symmetry, even though an accidental partial Klein symmetry of TM$_1$  mixing is predicted.
The simplicity of CSD3 as an origin of the reactor angle in the framework of indirect models contrasts
with the direct approach to the reactor angle which typically involves quite large family symmetry
groups. Alternatively one may appeal to charged lepton or higher order corrections
or else follow the semi-direct approach with only half the Klein symmetry preserved.
We have proposed an indirect $A_4\times Z_3^{10}$ family symmetry model which, at lowest order 
and without charged lepton corrections, explains the CSD3 
atmospheric flavon alignment $(0,1,1)$ and the solar flavon alignment $(1,3,1)$ via 
a mechanism based on F-term alignment, with $\eta = \mp \pi/3$ fixed by spontaneous CP violation.

In conclusion, we began in a model independent way by considering a two right-handed neutrino
model with a dominant texture zero, which has the feature that the number of physical observables
is equal to the number of see-saw parameters, which we related via a master formula.
Using the master formula, and the most recent oscillation data
we then searched for the simplest ratios of Yukawa couplings which can explain the data,
i.e. the minimal see-saw model 
with a normal neutrino mass hierarchy which can predict the PMNS parameters.  
Our quest was rewarded by the discovery of CSD3 which is remarkably simple and economical.
With only two input mass parameters, CSD3 describes nine physical observables, 
namely the PMNS mixing matrix plus the three neutrino masses. 
We have constructed an $A_4\times Z_3^{10}$ model based on CSD3, in which 
the input phase $\eta =\mp \pi /3$ is determined by spontaneous CP violation and 
the PMNS mixing matrix is completely fixed and only depends on 
the neutrino mass ratio $m_2/m_3$.  The model thus fixes all the mixing angles and phases
contained in the PMNS matrix and corresponds to 
approximate Tri-bimaximal-Cabibbo mixing, to an accuracy of one degree, with $\delta \approx \pm \pi/2$.
It is a candidate for the most minimal predictive
see-saw model consistent with current data
and can easily be tested by the forthcoming neutrino experiments.
In particular the prediction of maximal atmospheric mixing and a normal neutrino mass hierarchy
will be tested quite soon.

\section*{Acknowledgements}
SFK would like to thank A. Merle for help with MPT,
Christoph Luhn for very useful comments on the manuscript
and A. Kusenko and T. Yanagida and the IPMU for hospitality and support.
SFK also acknowledges partial support 
from the STFC Consolidated ST/J000396/1 and EU ITN grants UNILHC 237920 and INVISIBLES 289442 .


\begin{thebibliography}{99}

\bibitem{Minkowski:1977sc}
  P.~Minkowski,
  Phys.\ Lett.\  B {\bf 67} (1977) 421;
T. Yanagida, in Proceedings of theWorkshop on Unied Theory and Baryon Number
of the Universe, eds. O. Sawada and A. Sugamoto (KEK, 1979) p.95;
  P.~Ramond, 
Invited talk given at Conference: C79-02-25
(Feb 1979) p.265-280, CALT-68-709,
  hep-ph/9809459;
 M. Gell-Mann,
P. Ramond and R. Slansky, in Supergravity, eds. P. van Niewwenhuizen and D.
Freedman (North Holland, Amsterdam, 1979) Conf.Proc. C790927 p.315, PRINT-80-0576.

\bibitem{King:2004cx}
  S.~F.~King and T.~Yanagida,
  Prog.\ Theor.\ Phys.\  {\bf 114} (2006) 1035
  [hep-ph/0411030].
  
\bibitem{Davidson:2004wi}
  S.~Davidson,
  hep-ph/0409339.
    
\bibitem{Ade:2013lta}
  P.~A.~R.~Ade {\it et al.}  [Planck Collaboration],
  arXiv:1303.5076 [astro-ph.CO].
  
\bibitem{King:1998jw}
  S.~F.~King,
  Phys.\ Lett.\  B {\bf 439} (1998) 350
  [hep-ph/9806440];\,
  S.~F.~King,
  Nucl.\ Phys.\  B {\bf 562} (1999) 57
  [hep-ph/9904210];\,
  S.~F.~King,
  Nucl.\ Phys.\  B {\bf 576} (2000) 85
  [hep-ph/9912492];\,
  S.~F.~King,
  JHEP {\bf 0209} (2002) 011
  [hep-ph/0204360];
  T.~Blazek and S.~F.~King,
  Nucl.\ Phys.\ B {\bf 662} (2003) 359
  [hep-ph/0211368];
  S.~Antusch, S.~Boudjemaa and S.~F.~King,
  JHEP\ {\bf 1009} (2010) 096
  [arXiv:1003.5498].


  
  
\bibitem{King:1999mb}
  S.~F.~King,
  Nucl.\ Phys.\ B {\bf 576} (2000) 85
  [hep-ph/9912492];
  
\bibitem{King:2002nf}
  S.~F.~King,
  JHEP {\bf 0209} (2002) 011
  [hep-ph/0204360].

  
\bibitem{Frampton:2002qc}
  P.~H.~Frampton, S.~L.~Glashow and T.~Yanagida,
  Phys.\ Lett.\ B {\bf 548} (2002) 119
  [hep-ph/0208157].
  
  
\bibitem{Raidal:2002xf}
  M.~Raidal and A.~Strumia,
  Phys.\ Lett.\ B {\bf 553} (2003) 72
  [hep-ph/0210021].

  
\bibitem{King:2002qh}
  S.~F.~King,
  Phys.\ Rev.\ D {\bf 67} (2003) 113010
  [hep-ph/0211228].
  
   
\bibitem{Ibarra:2003up}
  A.~Ibarra and G.~G.~Ross,
  Phys.\ Lett.\ B {\bf 591} (2004) 285
  [hep-ph/0312138].

  
 
  

   
  

\bibitem{Gatto:1968ss}
  R.~Gatto, G.~Sartori and M.~Tonin,
  Phys.\ Lett.\ B {\bf 28} (1968) 128.

\bibitem{Georgi:1979df}
  H.~Georgi and C.~Jarlskog,
  Phys.\ Lett.\ B {\bf 86} (1979) 297.

\bibitem{An:2012eh}
  F.~P.~An {\it et al.}  [DAYA-BAY Collaboration],
  Phys.\ Rev.\ Lett.\  {\bf 108} (2012) 171803
  [arXiv:1203.1669];
   Y. Wang, talk at What is $\nu$? INVISIBLES'12 (Galileo Galilei Institute
   for Theoretical Physics, Florence, Italy, 2012); available at http://indico.cern.ch/conferenceTimeTable.py?confId=195985.
   
\bibitem{Ahn:2012nd}
  J.~K.~Ahn {\it et al.}  [RENO Collaboration],
  Phys.\ Rev.\ Lett.\  {\bf 108} (2012) 191802
  [arXiv:1204.0626].

\bibitem{Harrison:2002er}
  P.~F.~Harrison, D.~H.~Perkins and W.~G.~Scott,
  Phys.\ Lett.\ B {\bf 530} (2002) 167
  [hep-ph/0202074];

\bibitem{King:2007pr}
  S.~F.~King,
  Phys.\ Lett.\ B {\bf 659} (2008) 244
  [arXiv:0710.0530].
 

\bibitem{Pakvasa:2007zj}
  S.~Pakvasa, W.~Rodejohann and T.~J.~Weiler,
  Phys.\ Rev.\ Lett.\  {\bf 100} (2008) 111801
  [arXiv:0711.0052].

   
\bibitem{King:2012vj}
  S.~F.~King,
  Phys.\ Lett.\ B {\bf 718} (2012) 136
  [arXiv:1205.0506 [hep-ph]].
  
    
\bibitem{King:2013eh}
  S.~F.~King and C.~Luhn,
  arXiv:1301.1340 [hep-ph].
  
  \bibitem{SK}
  Talk by Y. Suzuki at Neutrino Telescopes XV, 11-15 March, 2013, Venice.
  
   \bibitem{T2K}
  Talk by T. Kobayashi at Neutrino Telescopes XV, 11-15 March, 2013, Venice, 
 
   

  
\bibitem{Harigaya:2012bw}
  K.~Harigaya, M.~Ibe and T.~T.~Yanagida,
  Phys.\ Rev.\ D {\bf 86} (2012) 013002
  [arXiv:1205.2198 [hep-ph]].


\bibitem{Shimizu:2012ry}
  Y.~Shimizu, R.~Takahashi and M.~Tanimoto,
  arXiv:1212.5913 [hep-ph].
  
  
 
\bibitem{King:2005bj}
  S.~F.~King,
  JHEP {\bf 0508} (2005) 105
  [hep-ph/0506297].


 
\bibitem{Chen:2009um}
  M.~-C.~Chen and S.~F.~King,
  JHEP {\bf 0906} (2009) 072
  [arXiv:0903.0125].
 


\bibitem{Choubey:2010vs}
  S.~Choubey, S.~F.~King and M.~Mitra,
  Phys.\ Rev.\ D {\bf 82} (2010) 033002
  [arXiv:1004.3756].

\bibitem{King:2009qt}
  S.~F.~King,
  Phys.\ Lett.\ B {\bf 675} (2009) 347
  [arXiv:0903.3199 [hep-ph]].
  
\bibitem{King:2011ab}
  S.~F.~King and C.~Luhn,
  JHEP {\bf 1203} (2012) 036
  [arXiv:1112.1959 [hep-ph]].
  
  
  
\bibitem{Antusch:2011ic}
  S.~Antusch, S.~F.~King, C.~Luhn and M.~Spinrath,
  Nucl.\ Phys.\ B {\bf 856} (2012) 328
  [arXiv:1108.4278 [hep-ph]].


\bibitem{Antusch:2013wn}
  S.~Antusch, S.~F.~King and M.~Spinrath,
  arXiv:1301.6764 [hep-ph].
  
\bibitem{Antusch:2011sx}
  S.~Antusch, S.~F.~King, C.~Luhn and M.~Spinrath,
  Nucl.\ Phys.\ B {\bf 850} (2011) 477
  [arXiv:1103.5930 [hep-ph]].
 
 
 
  
\bibitem{Antusch:2005gp}
  S.~Antusch, J.~Kersten, M.~Lindner, M.~Ratz and M.~A.~Schmidt,
  JHEP {\bf 0503} (2005) 024
  [hep-ph/0501272].
  
  
\bibitem{Boudjemaa:2008jf}
  S.~Boudjemaa and S.~F.~King,
  Phys.\ Rev.\ D {\bf 79} (2009) 033001
  [arXiv:0808.2782 [hep-ph]].
  

\bibitem{Antusch:2007ib}
  S.~Antusch, S.~F.~King and M.~Malinsky,
  Phys.\ Lett.\ B {\bf 671} (2009) 263
  [arXiv:0711.4727 [hep-ph]];
  S.~Antusch, S.~F.~King and M.~Malinsky,
  JHEP {\bf 0805} (2008) 066
  [arXiv:0712.3759 [hep-ph]];
  S.~Antusch, S.~F.~King and M.~Malinsky,
  Nucl.\ Phys.\ B {\bf 820} (2009) 32
  [arXiv:0810.3863 [hep-ph]].
  
\bibitem{Antusch:2010tf}
  S.~Antusch, S.~Boudjemaa and S.~F.~King,
  JHEP {\bf 1009} (2010) 096
  [arXiv:1003.5498 [hep-ph]].
  
  
\bibitem{Antusch:2006cw}
  S.~Antusch, S.~F.~King and A.~Riotto,
  JCAP {\bf 0611} (2006) 011
  [hep-ph/0609038].
  
\bibitem{Antusch:2011nz}
  S.~Antusch, P.~Di Bari, D.~A.~Jones and S.~F.~King,
  Phys.\ Rev.\ D {\bf 86} (2012) 023516
  [arXiv:1107.6002 [hep-ph]].


\bibitem{A4-refs}
  E.~Ma and G.~Rajasekaran,
  Phys.\ Rev.\  D {\bf 64} (2001) 113012
  [hep-ph/0106291];\,
  G.~Altarelli and F.~Feruglio,
  Nucl.\ Phys.\  B {\bf 720} (2005) 64
  [hep-ph/0504165];\,
  C.~Luhn, S.~Nasri and P.~Ramond,
  J.\ Math.\ Phys.\ \ {\bf 48} (2007) 073501
  [hep-th/0701188].



\bibitem{Lam:2006wm}
  C.~S.~Lam,
  Phys.\ Rev.\  D {\bf 74} (2006) 113004
  [hep-ph/0611017];
  C.~H.~Albright and W.~Rodejohann,
  Eur.\ Phys.\ J.\  C {\bf 62} (2009) 599
  [arXiv:0812.0436];
  C.~H.~Albright, A.~Dueck and W.~Rodejohann,
  Eur.\ Phys.\ J.\  C {\bf 70} (2010) 1099
  [arXiv:1004.2798];
  W.~Rodejohann and H.~Zhang,
  Phys.\ Rev.\ D {\bf 86} (2012) 093008
  [arXiv:1207.1225 [hep-ph]].
  
    
   
 
\end{thebibliography}
\end{document}